\documentclass[]{agujournal2019} 
\usepackage{url} 
\usepackage{lineno}
\usepackage[inline]{trackchanges} 
\usepackage{soul}
\usepackage{amsmath}
\usepackage{tabularray}

\journalname{Water Resources Research}

\begin{document}


\title{Physically-based dimensionless features for pluvial flood mapping with machine learning}






\authors{Mark S. Bartlett\affil{1}, Jared Van Blitterswyk\affil{2}, Martha Farella\affil{3}, Jinshu Li\affil{4}, Curtis Smith\affil{5}, Anthony J. Parolari\affil{6,7}, Lalitha Krishnamoorthy\affil{8}, Assaad Mrad\affil{9}}

\affiliation{1}{The Water Institute, Baton Rouge, LA, USA}
\affiliation{2}{Stantec, Ottawa, ON, Canada}
\affiliation{3}{Stantec, Flagstaff, AZ, USA}
\affiliation{4}{Stantec, Pasadena, CA, USA}
\affiliation{5}{Stantec, New York, NY, USA}
\affiliation{6}{Stantec, Lombard, IL, USA}
\affiliation{7}{Department of Civil, Construction, and Environmental Engineering, Marquette University, Milwaukee, WI, USA}
\affiliation{8}{Stantec, Austin, TX, USA}
\affiliation{9}{Stantec, Raleigh, NC, USA}


\correspondingauthor{Mark S. Bartlett}{Mark.Bartlett@gmail.com}

\begin{keypoints}
 \item Buckingham $\Pi$ theorem is used to create a dimensionless formulation of pluvial flood mapping.
\item A logistic regression machine learning model uses the dimensionless features to rapidly map pluvial flooding across geographies.
\item Dimensionless features outperform dimensional features, especially when the ML model is trained and tested in different regions.
\end{keypoints}

\begin{abstract}

Rapid delineation of flash flood extents is critical to mobilize emergency resources and to manage evacuations, thereby saving lives and property. Machine learning (ML) provides a promising solution for this rapid delineation, offering a computationally efficient alternative to high-resolution 2D flood models. However, even when trained on diverse geographic regions, ML models typically require retraining to perform well in new locations, and therefore often fail to generalize to never-before-seen conditions. To improve ML generalization, we apply Buckingham $\Pi$ theorem to derive dimensionless terms across multiple spatial scales. These multiscale $\Pi$ terms represent ratios of the relevant physical quantities governing the flooding process. Since the scaling laws of these dimensionless $\Pi$ terms encode process similarity across physical scales, these $\Pi$ terms enhance ML transferability to unseen locations. This is demonstrated by incorporating them as features in a logistic regression model for delineating flood extents. The features were calculated at different scales by varying accumulation thresholds for stream delineation. The ML flood maps, with an average AUC of 0.89, compared well with the results of 2D hydraulic models that are the basis of the Federal Emergency Management Agency flood hazard maps. The dimensionless $\Pi$ features outperformed dimensional features, with some of the largest gains in the AUC (of 20\%) occurring when the model was trained in one region and tested in another. Dimensionless and multi-scale $\Pi$  features in ML flood modeling have the potential to improve generalization, enabling mapping in unmapped areas and across a broader spectrum of landscapes, climates, and events.
\end{abstract}

\section*{Plain Language Summary}
Each year, flash floods are responsible for 80\%–90\% of all flood-related deaths, with approximately 40\% of these fatalities linked to pedestrian stream crossings or vehicles. Quickly predicting where flash floods will occur is crucial for saving lives and property. Traditional methods for predicting floods are slow and require significant computational resources, but machine learning (ML) can help speed up this process. However, existing ML models often have difficulty making accurate predictions for new areas with different climate patterns or landscape conditions. In this study, we improve ML flood prediction by representing the watershed with dimensionless numbers, which capture the relationships between key factors like rainfall and terrain shape, without relying on specific units (e.g., meters or seconds). By removing these units, the dimensionless numbers help ML models focus on the underlying patterns of flooding, making the ML more adaptable to different regions. We tested this approach with an ML model to predict flood risk, and the results showed that it performed well, even in regions where the ML had not been trained. This method could help create faster and more accurate flood maps for a wide range of areas, allowing communities to better prepare for floods.



\section{Introduction}

Floods cause dozens of deaths and billions of dollars in economic losses annually in the United States alone \cite{cornwall2021europe,ashley2008flood}. Urbanization and increases in extreme precipitation events due to climate change have already increased the frequency and severity of floods, which are anticipated to worsen with further climate change, making accurate and timely forecasting even more crucial \cite{wasko2021evidence}. To adequately mitigate flood impacts, rapid flood forecasting is needed, but this is still difficult to achieve in practice. Real-time flood forecasting at high resolution is partly prohibited by the high computational demand of high-resolution models \cite{ivanov2021breaking}.


Strategies to address the computational demand of high-resolution hydraulic models include inundation mapping, by filling digital terrain models (DTMs), and data-driven, or Machine Learning (ML), methods. Inundation mapping with high-resolution terrain data, such as the Height Above Nearest Drainage (HAND) procedure, is a promising approach to rapid flood mapping that can provide actionable information to first responders. The HAND procedure is an inundation mapping method that defines flood extent and depths based on the relative elevations above drainage points for a given water depth. This water depth can be derived from a flow forecast in conjunction with simplified hydraulic approaches such as Manning’s equation or rating curves. When using Manning’s equation, flood stage depth is determined from reach-averaged values of cross-sectional area, wetted perimeter, and flow surface area \cite{zheng2018river,zheng2018geoflood, scriven2021gis,garousi2019terrain}. Although these methods are fast, their limitations include a lack of bathymetry and infrastructure data and worse performance in smaller, low-order streams and small drainage pathways such as streets \cite{hocini2021performance,johnson2019integrated,garousi2019terrain}. Therefore, a key challenge in assessing pluvial flooding is accurately resolving flood risks from both small, low-order streams and larger, high-order streams. A location may be flooded by nearby small drainage pathways (e.g., streets) or by larger streams that inundate these smaller drainage pathways.

An alternative approach to improving the speed of flood forecasting is using machine learning (ML) to approximate hydraulic models for flood inundation mapping. Recently, ML approaches to flood mapping have received significant attention because of their speed and cost-effectiveness compared with previous approaches \cite{ivanov2021breaking,bentivoglio2022deep}. ML flood mapping models are trained against inundation extents from either satellite data, in-situ measurements, or hydrodynamic model output (e.g., HEC-RAS, LISFLOOD), and use a wide variety of input features  \cite{brunner2016hec,van2010lisflood,yang2021high}.
Typical input features include landscape and geomorphic characteristics, including elevation, curvature, topographic wetness index, slope, river density, drainage distance to the nearest river, HAND, and soil type \cite{bentivoglio2022deep,bui2018novel}. 
Such geomorphic and landscape features have been used as inputs to several linear binary classifiers \cite{nachappa2020flood, hosseiny2020framework, rahmati2020development, giovannettone2018statistical, collins2022predicting, samela2016based, manfreda2014flood, manfreda2014investigation, tavares2020predictive}. Features are generally selected based on data availability and hypotheses regarding their control of the physical processes governing runoff generation and flood development. 




One challenge facing the widespread adoption of ML flood models is their generalization to storms or locations outside of the training sample. Methods to improve model generalization can further alleviate constraints by limiting the data volume and computational time needed to apply models across large and diverse areas \cite{cache2024enhancing,pakdehi2023transferability,wagenaar2018regional,seleem2022transferability}. This challenge mirrors that of the over-parameterization of spatially-explicit, process-based models \cite{jakeman1993much,beven2001equifinality,beven2006manifesto}. One approach to address model over-parameterization and generalization is to transform the model into a reduced-dimension space that retains essential features of the underlying physics across catchments. For example, \citeA{hu2019rapid} apply linear decomposition to the input features and train the model in the reduced-order space. 

Alternatively, the Buckingham $\Pi$ theorem may be used to reformulate the governing equations of the physical process into a reduced-order, non-dimensional context  that captures the cause-effect relationship through a meta-level description, grounded in the physical similarity encoded in the scaling laws of the dimensionless $\Pi$ groups \cite{rudolph1998context, porporato2022hydrology, oppenheimer2023multi, gunaratnam2003improving}. Importantly, relative to linear decomposition, the Buckingham $\Pi$ approach constrains the number of dimensionless features based on the underlying physics. In unconstrained linear decomposition (e.g., principal component analysis or standard regression), the model often attempts to fit all available features, including those with little or no physical relevance. By constraining the model to fundamental dimensionless groups, the ML is more likely to capture general principles that apply across a wide range of similar systems, rather than fitting to irrelevant noise or features specific to certain conditions. Such an approach improves generalization, particularly when extrapolating across different physical scales (e.g.,  channel geometries, flow ratios) that are encountered when applying the ML model to a new watershed location. \cite{oppenheimer2023multi}. 

To evaluate the ability of this approach to improve ML flood model generalization, we apply the Buckingham $\Pi$ theorem to the underlying mass and momentum balance equations and develop a set of dimensionless hydraulic and hydrologic indices. In turn, these indices are used as inputs to a logistic regression model that predicts flooding across different geographies.   We hypothesize that applying the Buckingham  $\Pi$ theorem to derive dimensionless hydraulic and hydrologic indices will result in a more robust and generalizable logistic regression model for predicting flooding across different geographies. We test this hypothesis by training a logistic regression on these features for the Chicago area and applying the model to watersheds in northern New Jersey.

To address pluvial flooding from both small, low-order streams and larger, high-order streams, we define these dimensionless hydraulic and hydrologic indices relative to a point (i.e., a raster pixel), as well as in relation to two different drainage pathway delineations: local and non-local. The local delineation captures the small drainage pathways (e.g., streets) and low-order streams by using a low threshold for flow accumulation to determine stream origins. In contrast, the non-local delineation maps only higher-order streams by applying a much larger flow accumulation threshold. The logistic regression model then uses these dimensionless indices from the point scale and both the local and non-local delineations to predict flood risks across the various scales of drainage pathways.

\begin{figure}
\includegraphics[width=5.5 in]{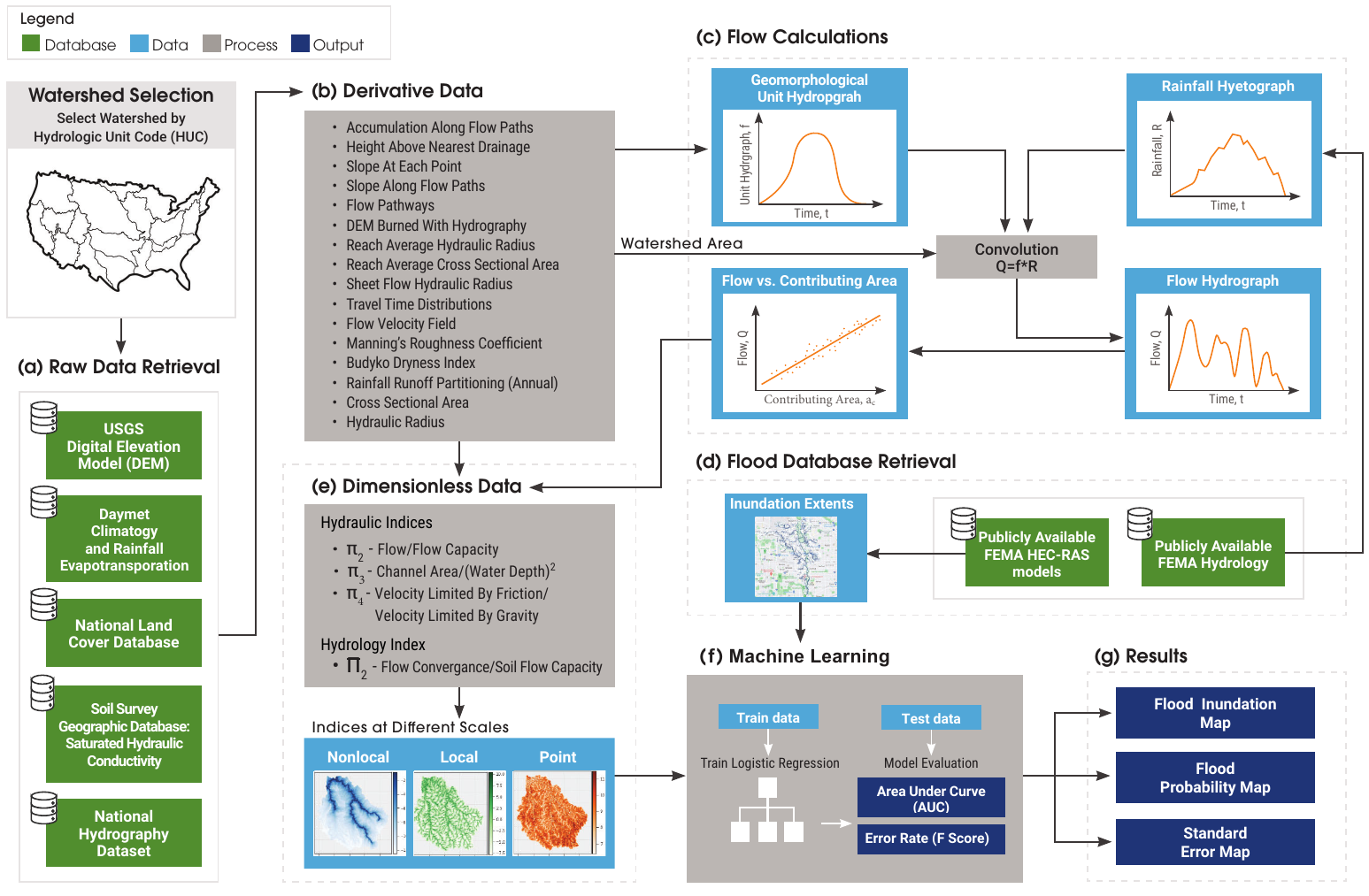}%
\caption{The ML process is applied to selected HUC 12 watersheds across the United States. For each selected HUC 12 watershed, a) the raw data is retrieved, b) the raw data is processed into derivative data, c) the derivative data and flood hydrology are used to construct a series of geomorphological instantaneous unit hydrographs to relate the peak (maximum) flow to contributing area d) the flood hydrology and extents are downloaded e) the flow and derivative data are combined into dimensionless indices that capture the hydraulic and hydrologic similarity of the flooding process, f) the dimensionless data is related to the flood extents by an ML logistic regression model, and g) the logistic regression model produces flood extents, flood probability, and the error in comparison to the original training data.  \label{fig:1_process}}
\end{figure}

The overall ML approach includes definition of the dimensionless input features, estimation of the input features, and the model training and testing (Fig. \ref{fig:1_process}). In Section \ref{sec:theory}, we use the Buckingham $\Pi$ theorem to derive dimensionless indices to capture the similarity of the flood process based on descriptions for both flow hydraulics and flood hydrology. We then discuss how the Buckingham $\Pi$ theorem-defined flood response function is subsumed by a logistic regression ML model. In Section \ref{sec:indices}, we discuss the steps used to process the dimensionless indices from raw data and the models used to approximate both flow and channel properties. To capture the flood process in detail, these dimensionless indices are defined at different scales: 1) at a point, 2) for smaller local flow paths (e.g., streets), and 3) for larger streams and rivers. A point is resolved at the finest resolution of the data, which is typically the resolution of the DEM. Section \ref{sec:ml} describes the ML methodology, including label data, model training, and performance assessment. In Section \ref{sec:application}, we compare the performance of the proposed dimensionless features with the traditional dimensional features, which are the derivative data of Fig. \ref{fig:1_process}.  This comparison is conducted with HUC 12 catchments in 2 distinct regions, Chicago and New Jersey. Finally we discuss the broad applicability of dimensionless indices in ML to improve generalization, interpretability, scalability, and efficiency.

\section{Theory}
\label{sec:theory}

Flooding is likely to occur in areas that are 1) adjacent to rivers, streams, and localized flow paths and 2) persistently wet (e.g., wetland areas or areas of flow convergence). The first case captures the storm event dynamics where the intensity of precipitation causes flow path expansion, while the second case captures the long-term controls of topography and climate on patterns of saturation. 
For each case, we consider the fundamental mathematical description of the flooding process, isolate the governing variables, and utilize the Buckingham $\Pi$ theorem to create dimensionless indices that relate the similarity of the flooding process across divergent conditions.

\begin{table}
\begin{minipage}{\textwidth}
\linespread{.5}\selectfont
\caption{Variable and parameter definitions$^a$ \label{tab:vars_params} }
\noindent
\begin{tabular}{l l l}
\hline
\noalign{\vskip 0.04in}
Symbol & Units & Description \\
\hline
\noalign{\vskip 0.05in}
$A$  & [L$^2$]  & Channel cross sectional area     \\
$A_s$ &   [L$^2$]    &Surface area of flooding \\
$a_c$  & [L$^2$]     &Contributing area  \\
$a_{c_r}$  & [L$^2$]    &Contributing area producing runoff  \\
$a_l$  & [L]     &Contributing area per unit contour length  \\
$D_I$  & [-]     & Budyko dryness index \\ 
$ET_{\max}$& [L/T]  &Long-term average potential evapotranspiration per unit area \\
$\langle ET \rangle$& [L/T]  &Long-term average evapotranspiration per unit area \\
$F_t$   & [-]    & Saturated fraction of the watershed \\
$f_{(\beta+(1 - \beta)F_t)}(t)$ & [1/T]     &Instantaneous unit hydrograph (IUH), i.e., travel time distribution \\
$g$    & [L/T$^2$]     & Gravitational acceleration     \\
$h$    & [L]     &Water depth (in channel) \\
$h_{\max}$ & [L] &Maximum water depth (in channel) during a storm event \\
$K$   &  [L$^3$/T]    & Conveyance  \\
$K_v$   & [L/T]    & Conveyance velocity \\
$k_s$    & [$L/T$]   &Saturated hydraulic conductivity \\
$l_c$  & [L]     &Channel length along the channel centerline \\
$n$   & [T/L$^{1/3}$]      &Manning's roughness coefficient  \\
$Q$  & [L$^3$/T]       &Flow rate in channel    \\
$Q_{max}$  & [L$^3$/T]       &Peak flow rate in channel    \\
$q_d$  &  [L$^2$/T]     &Groundwater flux  \\
$q_r$ &  [L$^2$/T]    &Recharge rate per unit contour length  \\
$q_o$  & [L$^2$/T]     &Subsurface flow per unit contour length \\
$r$   & [L/T]     & Recharge rate per unit area \\
$R_h$  & [L]     &Hydraulic radius  \\
$\langle R\rangle$& [L/T]& Long-term average rainfall intensity per unit area \\
$S$   & [-]      &Topographic slope at a point \\
$S_0$  & [-]     & Channel bed slope \\
$S_f$  & [-]     & Friction slope  \\
$t$   & [T]      & Time \\
$u$   & [L/T]      & Flow velocity\\
$\overline{w}$ &[L]&  reach averaged width  for a stream or river\\
$x$  & [L]       & Distance along the flow centerline  \\
$y_d$   & [L]    &Horizontal distance from nearest drainage \\
$z_w$   & [L]    &Distance to channel water \\
$z_d$   & [L]    & Vertical distance above nearest drainage \\
$\pi_1,...,\pi_n$ & [-] &$\Pi$-theorem terms related to hydraulics\\
$\Pi_1,...,\Pi_n$ & [-] &$\Pi$-theorem terms related to hydrology\\
\noalign{\vskip 0.04in}
\hline
\end{tabular}
\\
$^a$ Variables in the text with an overline bar indicate  averaged values, such as spatial averages (per unit area) or reach-averaged values as discussed in  \citeA{zheng2018river}.
\end{minipage}
\end{table}

\subsection{Storm Event Hydrology and Hydraulics}
\begin{figure}
\includegraphics[width=5.4 in]{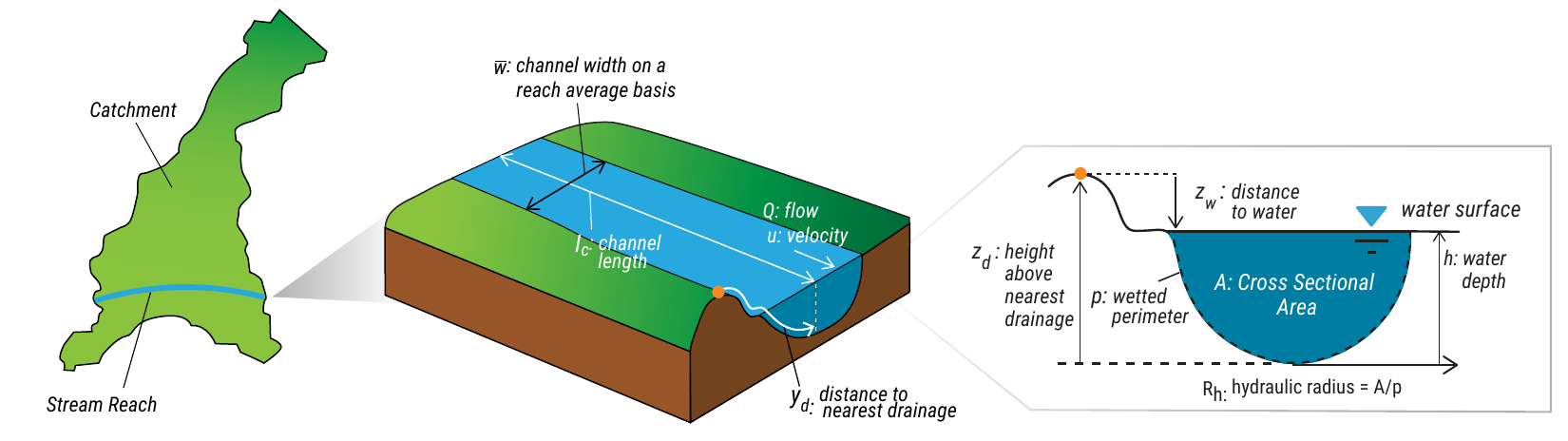}%
\caption{The governing parameters for the flow hydraulics, where at a point, flooding occurs when the distance to water is less than or equal to zero. \label{fig:1_flow_hydraulics}}
\end{figure}

For flooding by flow path expansion, the flow hydraulics are represented by the Saint–Venant equations describing conservation of mass (i.e., continuity) and momentum as follows \cite{chow1959open,moussa2000approximation, ercan2014scaling}:

\begin{equation}
\label{eq:continuity}
    \frac{\partial A}{\partial t} + \frac{\partial Q}{\partial x} = 0,  
\end{equation}
\begin{equation} 
\label{eq:momentum}
    \frac{\partial u_c}{\partial t} + u_c\frac{\partial u_c}{\partial x}+g\frac{\partial h}{\partial x}+g (S_f -S_0) = 0,
\end{equation}
where $t$ is time, $x$ is the spatial coordinate along the channel centerline, $h$ is the water depth, $Q$ is the flow rate, $A$ is the channel cross sectional area, $u_c$ is the depth averaged velocity in the $x$ direction, $g$ is gravitational acceleration, $S_0$ is the ground slope of the channel, and $S_f$ is the friction slope (Fig. \ref{fig:1_flow_hydraulics}). The Saint-Venant equations apply to a 1D topological description of the watershed flow path network.  Note that the flow, $Q$, could account for infiltration and exfiltration along the channel length, and in steady state, Eqs. (\ref{eq:continuity}) and (\ref{eq:momentum}) are the basis of the standard step method of HEC-RAS 1D. The coupled set of equations (\ref{eq:continuity}) and (\ref{eq:momentum}) is closed by assuming a function for the friction slope \cite{yang2019lattice, ercan2014scaling}, i.e.,

\begin{equation}
    S_f = \frac{Q|Q|}{K^2},
\end{equation}
where $K = \frac{1}{n}R_h^{2/3}A$ is the conveyance, for which $R_h$ is the hydraulic radius, and $n$ is Manning's roughness coefficient.

For a flood event, equations (\ref{eq:continuity}) and (\ref{eq:momentum}) determine the dynamics of water flow and the hydraulic pressure and depths, where $h_{max}$ is the maximum depth at a point along the 1D topological description of the channel. We assume that any given point is flooded when $h_{\max}$ exceeds the elevation above the nearest drainage, $z_d$ (Fig. \ref{fig:1_flow_hydraulics}). That is, a point is flooded when the distance to water is negative, $z_w =  z_d-h_{\max} \leq 0$.
We assume $z_w$ depends on: 
\begin{enumerate}
\setlength{\itemsep}{0pt}
   \item $Q$ [L$^3$/T], the flow rate, for the flood magnitude constraint,
   \item $z_d$ [L], the height above the nearest drainage, for constraining the number of flooded points,
   \item $K_v$ [L/T], the velocity of conveyance (i.e., $K/A$) of flow representing the channel capacity constraint,
   \item $A$ [L$^2$], the channel cross sectional area,
   \item $g S_0 $ [L/T$^2$], the slope multiplied by gravity, representing the external force driving the flow downslope.
\end{enumerate}
The velocity, $u_c$, and the friction slope, $S_f$, are not considered since they are derived directly from the listed variables. Furthermore, we do not consider time because flooding is based on the maximum flow and height of the water level, $Q_{\max}$ and $h_{\max}$, respectively. Hence, the depth of the water, $z_w$, is considered the maximum level of water during a flood event.

Based on the Saint-Venant equations and the assumed channel geometry, the distance to water is a function of 5 variables, that is, $z_w$ = $f(Q, K_v, z_d, A, g S_0)$, and two dimensions of length [L] and time [T]. Following the Buckingham $\Pi$ theorem, the required number of $\pi$ groups is $6-2= 4$, where we have 1 dependent variable and 5 independent variables. The $\pi$ groups based on the two repeating variables, $K_v$ and $A$, are

\begin{align}
    \label{eq:pi_1}
    \pi_1 &= \frac{z_w}{\sqrt{A}}, \\
    \label{eq:pi_2}
    \pi_2 &= \frac{Q}{K_v A}, \\
    \pi_3 &= \frac{z_d}{\sqrt{A}}, \\
    \label{eq:pi_4}
    \pi_4 &= \frac{g S_0 \sqrt{A}}{K_v^2},
\end{align}
where the groupings were selected by inspection. This process yields the following dimensionless $\pi$ group relationship

\begin{align}
    \pi_1 = f(\pi_2, \pi_3, \pi_4),
\end{align}
where $\pi_1$, the distance to the flood surface, $z_w$, normalized by the square root of the channel cross sectional area, $\sqrt{A}$, is a function of the other terms, $\pi_2$, $\pi_3$, and $\pi_4$. The term $\pi_2$ represents the flow to flow capacity ratio, $\pi_3$ accounts for how the geometry of the channel (e.g., expansion and contraction of the channel) impacts flooding, while $\pi_4$ represents the flow velocity constrained by the gravitational force over the flow velocity constrained by the frictional force. Note that $\pi_3$  does not exist without reference to a stream or river to define the height above the nearest drainage, $z_d$.

\subsection{Long-term Hydrology}
\begin{figure}
\includegraphics[width=5.4 in]{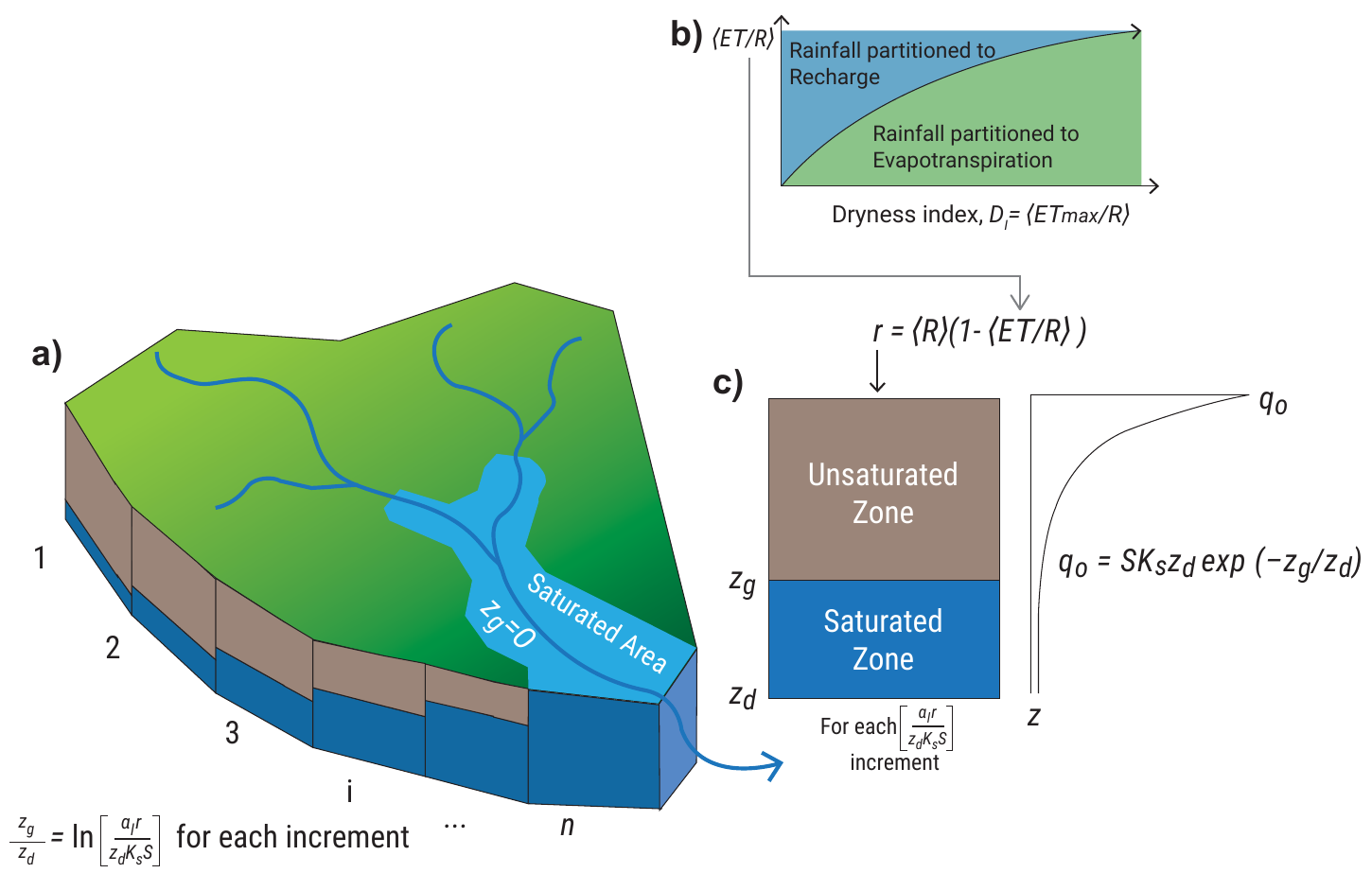}%
\caption{The watershed hydrology for calculating wet saturated areas (prone to flooding) is based on a)  the TOPMODEL assumption of a spatially variable inflow rate derived from the contributing area (per unit contour length), $a_l$,  multiplied by the watershed recharge rate, $r$, here derived from b) a partition of rainfall to evapotranspiration, $\langle ET/R\rangle$, given by the Budyko curve, that is balanced by  c) a water table that parallels the surface slope with soil water lateral transmission, $q_o$, that decreases exponentially with depth. \label{fig:2_TOPModel}}
\end{figure}

To identify flooding in typically wet areas with saturated soil, we consider the assumptions of TOPMODEL \cite{beven2012rainfall} where the dynamics of the water table depth from the surface are considered as a secession of steady states that balance the long-term watershed recharge with the outflow from the saturated soil layer. Since fluxes are assumed to be perpendicular to the downslope gradient, they are expressed on a per unit contour length basis, which simplifies the representation of water movement across the landscape \cite{beven2012rainfall}. Accordingly, the long term recharge (per unit contour length), $q_r$, is given by
 
\begin{align}
    q_r &= a_l r,
    \label{eq:qr_recharge}
\end{align}
where $a_l$ is the contributing area (per unit contour length), and $r$ is the uniform recharge rate (per unit area). Note that $a_l$ is equal to the contributing area, $a_c$.
In steady state,  $q_r$ is balanced by the outflow governed by Darcy's Law with the hydraulic gradient, $dh/dx$, approximated by the local topographic slope, $S$ \cite{beven2012rainfall}, and a hydraulic conductivity that decreases exponentially with the soil depth, i.e.,

\begin{align}
    k_s \int_{z_g}^{\infty}e^{-\frac{z}{z_d}}dz &= k_s z_d e^{-\frac{z_g}{z_d}},
\end{align}
where $k_s$ is the saturated hydraulic conductivity at the surface that decreases exponentially with a decay factor of $1/z_d$ \cite{ducharne2009reducing}, $z_g$ is the depth to groundwater, and $z$ is the depth from the surface (Fig. \ref{fig:2_TOPModel}). Unlike TOPMODEL, we have assumed that the decay factor is the reciprocal of the HAND value at each point. Consequently, the average hydraulic conductivity occurs at the elevation of the nearest drainage flow path. Accordingly, the water table outflow (per unit contour length) is

\begin{align}
    q_o &=   S k_s z_d e^{-\frac{z_g}{z_d}},
    \label{eq:qo_outflow}
\end{align}
where $q_0$ is the (per unit contour length) outflow. 


In steady state, the inflow of Eq. (\ref{eq:qr_recharge}) equals the outflow of Eq. (\ref{eq:qo_outflow}) and, after rearranging terms, we retrieve the relationship between two dimensionless terms,

\begin{align}
    \frac{z_g}{z_d}= - \ln \left[\frac{a_l \ r}{z_d k_s S}\right],
    \label{eq:3_z_g}
\end{align}
where the depth to groundwater, $z_g$, normalized by the height above drainage, $z_d$, is equivalent to the natural log of the recharge inflow normalized by the outflow of Darcy's Law with the hydraulic gradient subsumed by the topographic slope, $S$, over a depth equal to the height above nearest drainage (per unit contour length) (Fig. \ref{fig:2_TOPModel}).

Based on Eq. (\ref{eq:3_z_g}), the distance to the surface of the groundwater, $z_g = f(z_d, q_r,q_d)$, depends on

\begin{enumerate}
\setlength{\itemsep}{0pt}
   \item $z_d$ [L], the height above nearest drainage, for constraining the number of flooded points,
   \item $q_r$ [L$^2$/T], the moisture recharge flux into a point per unit contour length,
   \item $q_d$ [L/T], the groundwater flux based on Darcy's Law, i.e.,  $k_s S$, assuming the hydraulic gradient, $dh/dx$, is represented by the topographic surface slope, $S$.
\end{enumerate}
Therefore, the system consists of 4 variables and 2 dimensions, so the Buckingham $\Pi$ theorem yields two $\Pi$ groups ($4-2 = 2$). From Eq. (\ref{eq:3_z_g}), these $\Pi$ groups are

\begin{align}
    \Pi_1 &= \frac{z_g}{z_d}, \\
    \Pi_2 &= \frac{a_l \ r}{z_d \  k_s \ S},
    \label{eq:Pi_2}
\end{align}
where we have substituted $q_r$ with $a_l r$ from Eq. (\ref{eq:qr_recharge}) and $q_d$ with $k_s S$ following the assumption of TOPMODEL. These $\Pi$ groups provide for the following dimensionless relationship

\begin{align}
    \Pi_1 = \ln(\Pi_2),
    \label{eq:pi_func_hydro}
\end{align}
where the natural log function is inferred from inspection of Eq. (\ref{eq:3_z_g}). The term $\ln(\Pi_2)$ is similar to the topographic wetness index \cite{beven1979physically}; but, differently, the argument of the natural log function, $\Pi_2$, is dimensionless and reflects the ratio of water accumulation to water outflow from groundwater (per unit countour length). In $\ln(\Pi_2)$, the addition of $z_d$ is similar to the topographic wetness index (TWI) that accounts for the landscape position relative to water (groundwater and surface water), which was found to better represent field observations \cite{meles2020wetness}. 

\subsection{From Buckingham $\Pi$ theorem to ML}

Following the Buckingham $\Pi$ theorem, flooding is determined by the dimensionless groups $\pi_1$ or $\Pi_1$ for flooding from expansion of the channel flow path and in saturated areas (i.e., lakes and depressions), respectively. In both cases, when $\pi_1$ or $\Pi_1$ is negative, an area is flooded, but when both $\pi_1$ and $\Pi_1$ are positive, an area is not flooded. Thus, the binary result of flooded (e.g., 1) and not flooded (e.g., 0) is found by applying a multidimensional Heaviside function, $\Theta(\cdot, \cdot)$, i.e., 

\begin{align}
  1-\Theta(\Pi_1,\pi_1) = 1-\Theta[f(\pi_2, \pi_3, \pi_4) ,  \ln(\Pi_2)],
  \label{eq:pi_func_final}
\end{align}
for which the Heaviside step function is only 1 if both arguments are positive, the function is right continuous, i.e., $\Theta(0, 0) = 1$, and the function  $f(\pi_2, \pi_3, \pi_4)$ is unknown. Here, we consider that an ML algorithm may subsume the overall function $\Theta(f(\pi_2, \pi_3, \pi_4), \ln(\Pi_2))$. Specifically, we consider the logistic regression for the likelihood (probability of flooding), i.e.,

\begin{align}
    1-\Theta(\Pi_1 , \pi_1) \approx \Theta[P(\ln[\pi_2], \ln[\pi_3], \ln[\pi_4],\ln[\Pi_2])-\epsilon)],
    \label{eq:pi_func_ml}
\end{align}
where $P(\ln[\pi_2], \ln[\pi_3], \ln[\pi_4],\ln[\Pi_2])$ is the logistic regression probability, the natural log, $\ln[\cdot]$, is applied to all terms because we assume a multiplicative relationship (i.e., addition in log space results in multiplication of the variables), and the discrimination threshold, $\epsilon$, allows for the step function, $\Theta(\cdot)$, to filter the logistic regression probability into flooded (i.e., 1) or not flooded areas (i.e., 0). 

\begin{table}
\begin{minipage}{\textwidth}
\small
\linespread{.5}\selectfont
\caption{Input data sources and resolution \label{tab:data}}
\noindent
\begin{tabular}{l l p{4cm} p{5cm}}
\hline
\noalign{\vskip 0.04in}
Data Type & Resolution (m) & Source & URL\\
\hline
\noalign{\vskip 0.05in}
Digital Elevation Model & 1-10 & USGS Elevation Products (3DEP)  & https://prd-tnm.s3.amazonaws \allowbreak.com/index.html?prefix=\allowbreak StagedProducts/Elevation/  \\
Land Use & 30 & NLCD$^a$ & https://www.mrlc.gov/data \\ 
Impervious Surface & 30 & NLCD Impervious Surface            & https://www.mrlc.gov/data\\ 
Roads & 30 & NLCD Impervious \allowbreak Descriptor & https://www.mrlc.gov/data \\ 
Soil Type          & 30  & NRCS SSURGO$^b$        &  https://nrcs.app.box.com/v/\allowbreak soils/folder/17971946225  \\
Hydrography        & n/a & National Hydrography Dataset (NHD) &  https://prd-tnm.s3.amazonaws \allowbreak.com/index.html?prefix=\allowbreak StagedProducts/Hydrography/  \\
Watershed Boundaries & n/a & USGS Hydrologic Units (HU)       & https://prd-tnm.s3.amazonaws \allowbreak.com/index.html?prefix=\allowbreak StagedProducts/Hydrography/\\
Flood Maps        & 1-5  & FEMA Risk MAP Program             & https://msc.fema.gov \\
Climatology    & 10$^3$ & Daymet                            & https://thredds.daac.ornl.gov/ \allowbreak thredds/catalog/ornldaac/ \allowbreak 2130/catalog.html \\
\noalign{\vskip 0.04in}
\hline
\end{tabular}
\\
$^a$ National Land Cover Database \\
$^b$ Natural Resources Conservation Service Soil Survey Geographic Database
\end{minipage}
\end{table}

\section{Data and Machine Learning Methodology}
\label{sec:data_methods}

\subsection{Dimensionless Input Feature Estimation}
\label{sec:indices}

The input feature $\pi$ and $\Pi$ groups were derived from the variables of $Q$, $A$, $n$, $R_h$, $z_d$, $S_0$, $a_l$, $r$, $k_s$, and $S$ (Table \ref{tab:vars_params}). For each HUC-12 watershed, these variables were derived by transforming the raw data (Table \ref{tab:data}) using GRASS GIS \cite{jasiewicz2011new}. Before processing the variables with GRASS GIS, the raw DEM was hydro-enforced with the stream and river hydrography centerlines \cite{USGS_NHD}. Here, the raw DEM refers to the standard DEM product from the USGS 3D Elevation Program (3DEP), which follows a consistent set of specifications and is typically available at 1-m, 3-m, or 10-m resolutions \cite{USGS_3DEP_DEM}. The resolution of the DEM does not significantly affect the calculated values; rather, a lower-resolution DEM results in coarser spatial detail without altering the overall estimates of the input features. The DEM with the highest available resolution was used to generate the $\pi$ and $\Pi$ groups at different hydrographic scales.

\subsubsection{Landscape properties at each point}
\label{sec:point_properties}

Landscape properties at each point were calculated on a grid at the resolution of the raw DEM data, including Manning's $n$, saturated hydraulic conductivity, slope, contributing area, conveyance velocity, and recharge rate. Manning's roughness coefficient, $n$, was based on land use data as described in the Hydrologic Engineering Center's River Analysis System (HEC-RAS)  \cite{brunner2016hec,USGS_2024_NLCD}. Saturated hydraulic conductivity, $k_s$, was extracted from the Soil Survey Geographic Database (SSURGO) database as the most restrictive soil component (i.e., lowest $k_s$ value) in the top 5 meters \cite{SSURGO_Database}. Slope and contributing area were estimated from the DEM. Based on the hydro-enforced DEM, the contributing area, $a_c$, was derived from the flow accumulation output of the GRASS GIS \emph{r.watershed} module (representing the number of raster cells contributing to each cell via flow direction), multiplied by the raster cell area (determined by the DEM resolution). This approach resulted in an $a_c$  independent of the DEM resolution.


The conveyance velocity, \(K_v\), was calculated in two steps: 
1) initially, we estimated \(K_v\) based on an assumed sheet flow and Manning's n, and 
2) we then replaced these initial point values with an average of the upstream conveyance velocities (tributary to each point), scaled by the factor \(\frac{a_c^{1/2}}{\overline{a_c^{1/2} S^{1/2}}}\), where the overline (of $\overline{a_c^{1/2} S^{1/2}}$) indicates the upstream average of all contributing points \cite{rinaldo1991geomorphological, odorico2003hillslope, maidment1996unit}.  For the initial calculation of the point conveyance velocity, we assumed a hydraulic radius corresponding to a sheet flow condition with a depth of 5 cm—a value that should be linked to the rainfall intensity in future work. This initial assumption of sheet flow provided a practical and consistent means to estimate the hydraulic radius in Manning’s equation, offering a uniform basis for determining conveyance velocity across the watershed. While this approximation may require refinement, it established a baseline for velocity estimation. As flow accumulated and transitioned from diffuse overland flow into well-defined channels, the scaling factor $\frac{a_c^{1/2}}{\overline{a_c^{1/2} S^{1/2}}}$ effectively adjusted the hydraulic radius to account for the progressive deepening of the flow and the corresponding reduction in the effective resistance to the flow exerted by the bed and banks \cite{maidment1996unit}. This adjustment allowed the estimated velocities to reflect the evolving hydrodynamic conditions, capturing the transition from shallow sheet flow in headwaters to concentrated channelized flow downstream.


At each point, the long-term recharge rate, $r$, was approximated as $r = \langle R\rangle(1 - \langle ET/R\rangle)$, where $\langle R \rangle$ represents the average  rainfall intensity on an daily average basis, and $\langle ET/R \rangle$ is the average evapotranspiration-to-rainfall ratio. This ratio $\langle ET/R\rangle$ was retreived from the Budkyo curve equation $\langle  ET/R \rangle = \{ D_I [1-\exp(-D_I)]\tanh(1/D_I)\}^{0.5}$, where $D_I = ET_{\max}/\langle R \rangle$ was the dryness index based on the average potential evapotranspiration on a daily basis,  $ET_{\max}$ \cite{rodrigueziturbe2004ecohydrology, porporato2022ecohydrology}. These values of $ET_{\max}$ and $\langle R \rangle$, both on a daily average basis, were obtained from the Daymet dataset at a 1-km resolution for 2022 \cite{thornton2022daymet}.




\subsubsection{Peak flow model}

To determine the maximum flood extents, we replaced the time varying flow, $Q(t)$, of $\pi_2$ with a peak flow $Q_{\max}$ taken from a flow hydrograph specific to each storm event (Fig. \ref{fig:1_process}). This flow hydrograph captured the spatial runoff variability based on a semi-distributed rainfall-runoff model \cite{bartlett2015unified2}. In this approach, the watershed was conceptualized as having two distinct hydrologic connections to the stream:
\begin{itemize}
\item Over a fraction of the area, $\beta$, runoff is generated before soil saturation. This fraction may represent riparian zones and lower hillslopes that maintain a persistent connection to the stream. In urban settings, $\beta$ also may include streets and impervious surfaces engineered for direct drainage to streams and rivers.
\item Over the remaining area, $1-\beta$, runoff occurs only after soil saturation is reached.
\end{itemize}
When also considering the saturated fraction of the watershed, $F_t$, the unit-area runoff, $\overline{Q}$, varied across three distinct areas:

\begin{align}
    \label{eq:Qbeta}
      &\overline{Q}_{\beta} = \overline{R} F_t(\overline{S}, \overline{R})+ \overline{R}(1-F_t(\overline{S}, \overline{R}))(1-\overline{c}),\\
      &\overline{Q}_{(1-\beta)F_t} = \overline{R} \label{eq:Q(1-beta)}, \\
      &\overline{Q}_{(1-\beta)(1-F_t)} = 0,
\end{align}
where $\overline{R}$ is unit area rainfall, and
over the fraction of area, $(1-\beta)$, runoff only occurs over the saturated fraction of area, $F_t$ \cite{bartlett2015unified1,bartlett2015unified2,bartlett2017reply}. Note that the overall unit area runoff, $\beta \overline{Q}_{\beta} +(1-\beta)F_t \overline{Q}_{(1-\beta)F_t}$, is equal to the general runoff equation (10) of \citeA{bartlett2015unified2}. The fraction of saturated area, $F_t$, is calculated based on a hydrology model, e..g., the variable infiltration capacity (VIC) model, TOPMODEL, and the NRCS-CN approach \cite{bartlett2015unified1,bartlett2015unified2,bartlett2017reply}. 

For this study, the fraction of area $\beta$ was reasonably represented by streets and other impervious areas. In these  areas, it was assumed that the antecedent moisture deficit, $\overline{c}$, was negligible, i.e., $\overline{c}\approx 0$. Thus, within the fraction of area $\beta + (1-\beta)F_t$, the runoff was simply $\overline{R}$. With these assumptions, the peak flow was calculated as,
 
\begin{align}
    Q_{max} =& \max\left[a_{c_r} \int_0^t f_{(\beta+(1 - \beta)F_t)}(t-\tau)\overline{R}_t(\tau)d\tau \right],
    \label{eq:Qmax_simple}
\end{align}
where  $f_{(\beta+(1-\beta)F_t)}(\cdot)$ is the geomorphological instantaneous unit hydrograph (GIUH) for the fraction of watershed area, $\beta+(1-\beta)F_t$, where the unit area runoff is the unit area rainfall, $\overline{R}=\int_0^{\infty}\overline{R}_t(\tau)d\tau$, for which $\overline{R}_t(\tau)$ is the rainfall hyetograph, $a_{c_r}$ is the runoff producing area, i.e.,  $\beta+(1-\beta)F_t$ multiplied by the watershed area, and $\tau$ is the time from the beginning of the storm.  For Eq. (\ref{eq:Qmax_simple}), the GIUH, $f_{(\beta+(1-\beta)F_t)}(\cdot)$, was based on the travel times of the watershed points within the fraction of area $\beta + (1-\beta)F_t$. At each point, travel times were computed based on the distance (following flow direction) to the watershed outlet, which was determined using the GRASS GIS function  \emph{r.stream.distance} \cite{jasiewicz2011new}. This total distance was decomposed into two components: a hillslope distance, $x_h$, and a channel distance, $x_c$. Corresponding average water velocities in these areas, $\overline{u}_h$ for the hillslope and $\overline{u}_c$ for the channel, respectively were the product of the spatially averaged slope and the conveyance velocity, $\overline{K_v} \: \overline{S^{0.5}}$. In these areas, the conveyance velocity at each raster pixel was calculated as described in Section \ref{sec:point_properties}. The travel time at each point then was calculated as $t = \frac{x_h}{\overline{u}_h} + \frac{x_c}{\overline{u}_c}$ \cite{rigon2011geomorphic}.

The GIUH was constructed by considering the travel times over the watershed points within two areas: the impervious area fraction $\beta$ and the saturated area fraction $(1-\beta)F_t$. For the impervious area fraction $\beta$, the points considered were the streets and other impervious areas, as derived from the NLCD impervious descriptor map \cite{USGS_2024_NLCD}. For the remaining pervious area, the saturated points $(1-\beta)F_t$ were determined using $\Pi_2$, which is a wetness index akin to the TWI  \cite{meles2020wetness}. Specifically, the saturated points representing $(1-\beta)F_t$ were identified as those locations where $\Pi_2$ exceeded the value corresponding to the boundary of the unsaturated area, i.e.,

\begin{align}
\Pi_2 \geq P^{-1}_{\Pi_2}(1-F_t),
\end{align}
in which the boundary value is given by empirical quantile function $P^{-1}_{\Pi_2}(1-F_t)$ where the argument is the fraction of unsaturated area $1-F_t$. This empirical quantile function, $P^{-1}_{\Pi_2}(\cdot)$, was constructed from a rank ordering of the $\Pi_2$ values over the fraction of area $1-\beta$. The fraction of saturated area, $F_t$, was calculated with Eq. (24) of \citeA{bartlett2015unified1} for the extended NRCS-CN method, which in this case (for $\overline{c}\approx 0$) was 

\begin{align}
F_t = \frac{\overline{R}(1-\beta)}{\overline{S}+\overline{R}(1-\beta)},
\label{eq:NRCS_CNx}
\end{align}
where the maximum potential retention, $\overline{S}$, is based on the curve number (CN).

\begin{figure}
\includegraphics[width=5 in]{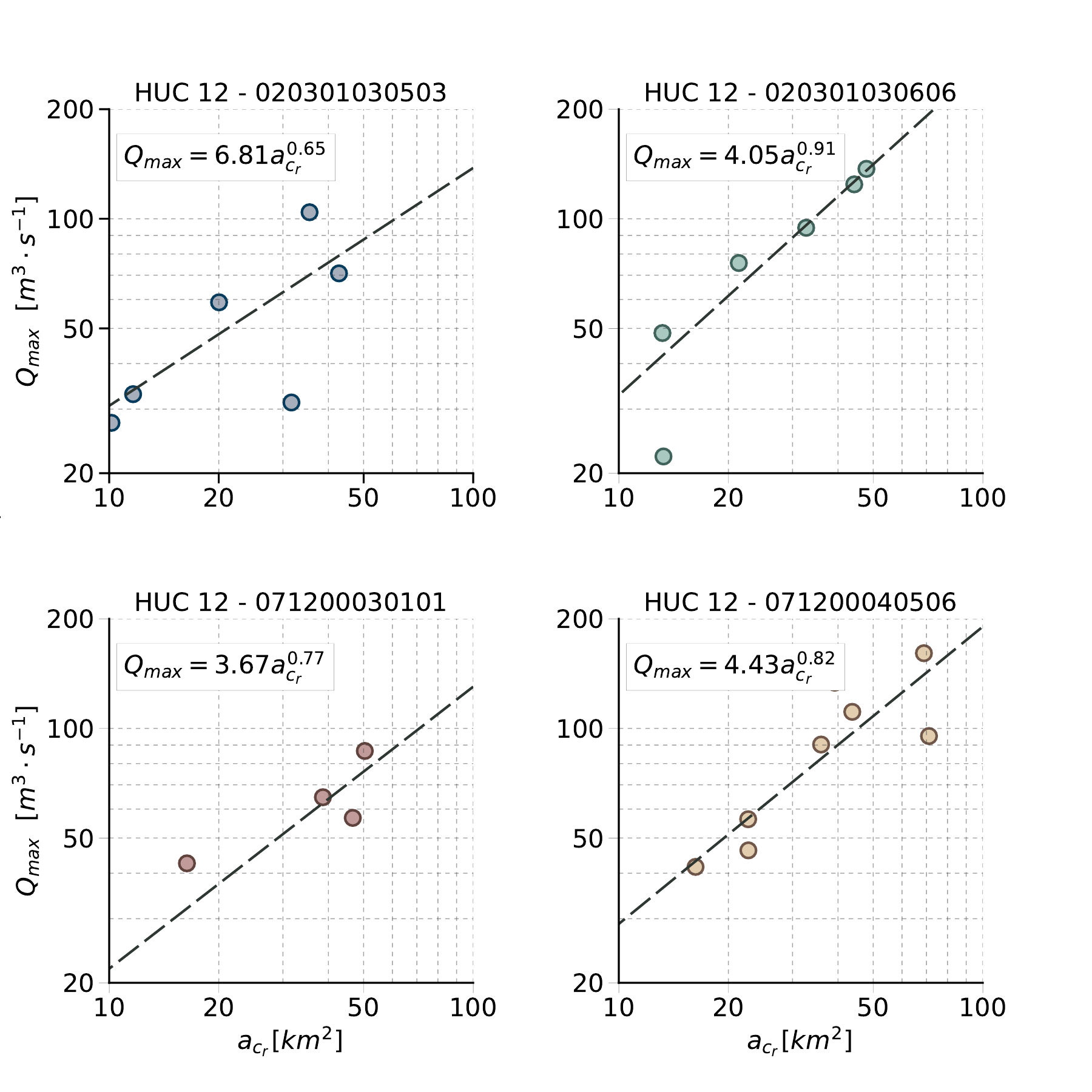}%
\caption{The relationship between peak flow, $Q_{max}$, and the runoff producing contributing area, $a_{c_r}$, is based on a series of $Q_{\max}$ values calculated with Eq. (\ref{eq:Qmax_simple}). Each $Q_{\max}$ is derived from 1) a GIUH specific to the runoff producing area, $a_{c_r}$, and 2) a rainfall hyetograph based on the 100-year rainfall event (Table S1) and the 2nd quartile, 50th decile event temporal distribution from NOAA Atlas 14. \label{fig:5_flow_vs_area}}
\end{figure}

For any unit-area precipitation hyetograph, the peak flow was defined through Eq. (\ref{eq:Qmax_simple}). In turn, the maximum flow roughly was inferred as a function of the contributing area \cite{nardi2006investigating,rigon2011geomorphic}, i.e.,

\begin{align}
     Q_{max} = c_1 \left(a_{c_r}\right)^{c_2},
     \label{eq:Qmax(Aw)}
\end{align}
where the contributing area $a_{c_r}$ was derived only accounting for the area producing runoff, i.e., the fraction of watershed area $\beta + (1-\beta)F_t$, and $c_1$ and $c_2$ were found by fitting the equation to multiple (watershed specific) data points derived from the area and flow relationship of Eq. (\ref{eq:Qmax_simple})  (Fig. \ref{fig:5_flow_vs_area}). 



Consequently, for each HUC 12 watershed, we divided the watershed into a series of nested watersheds down to an area of approximately 10 km$^2$, and calculated a series of data points that related the contributing area to the maximum flow. Subsequently, we found $c_1$ and $c_2$ from the best-fitting line and then mapped the contributing area to the peak flow based on Eq. (\ref{eq:Qmax(Aw)}) \cite{rigon2011geomorphic}. Generally, a log-linear relationship adequately described how the peak flow scaled with the contributing area for each HUC 12 watershed (Fig. \ref{fig:5_flow_vs_area}). Precipitation inputs to the watersheds were estimated to be consistent with the FEMA HEC-RAS approach (Table S2). The precipitation hyetographs, $\overline{R}_t(\tau)$, for the 10-, 100-, and 1000-yr. events were estimated from the NOAA Atlas 14 24-hour event for both the total rainfall amount and the temporal distribution, which was the second quartile, 50th decile event \cite{bonnin2004atlas}. This precipitation hyetograph was then input into the $Q_{max}$ flow model of Eq. (\ref{eq:Qmax_simple}) with the saturated fraction of area of Eq. (\ref{eq:NRCS_CNx}) based on the maximum potential retention $\overline{S}$ derived from the CNs of Table S2.

\subsubsection{Channel properties}
The channel network was delineated based on a specified flow accumulation threshold, culminating in the assignment of a flow capacity and hydraulic parameters at each watershed point. Channel delineation thresholds for local and non-local scales were 0.01 mi$^2$ and 1 mi$^2$, respectively. For the respective channel networks, each watershed point was related to a channel flow capacity when the channel stage equaled the HAND of the point. This capacity was based on $n$ and $S_o$ from the nearest downslope channel centerline  and $A$ and $R_h$ respectively calculated as the product of the reach averaged hydraulic values (at the HAND depth) and an adjustment factor based on the channel geometry specific to the point (\ref{sec:App_Rh}). Note that the reach averaged hydraulic values were calculated based on the overall, respective nonlocal and local channel networks \cite{zheng2018geoflood,garousi2019terrain}. For channel networks extracted from the DEM, the inclusion of bathymetry would alter the extracted flow paths and provide a more accurate representation of the channel centerlines and channel geometry.

\subsubsection{Indices at different scales}

\begin{figure}
\includegraphics[width=5.5 in]{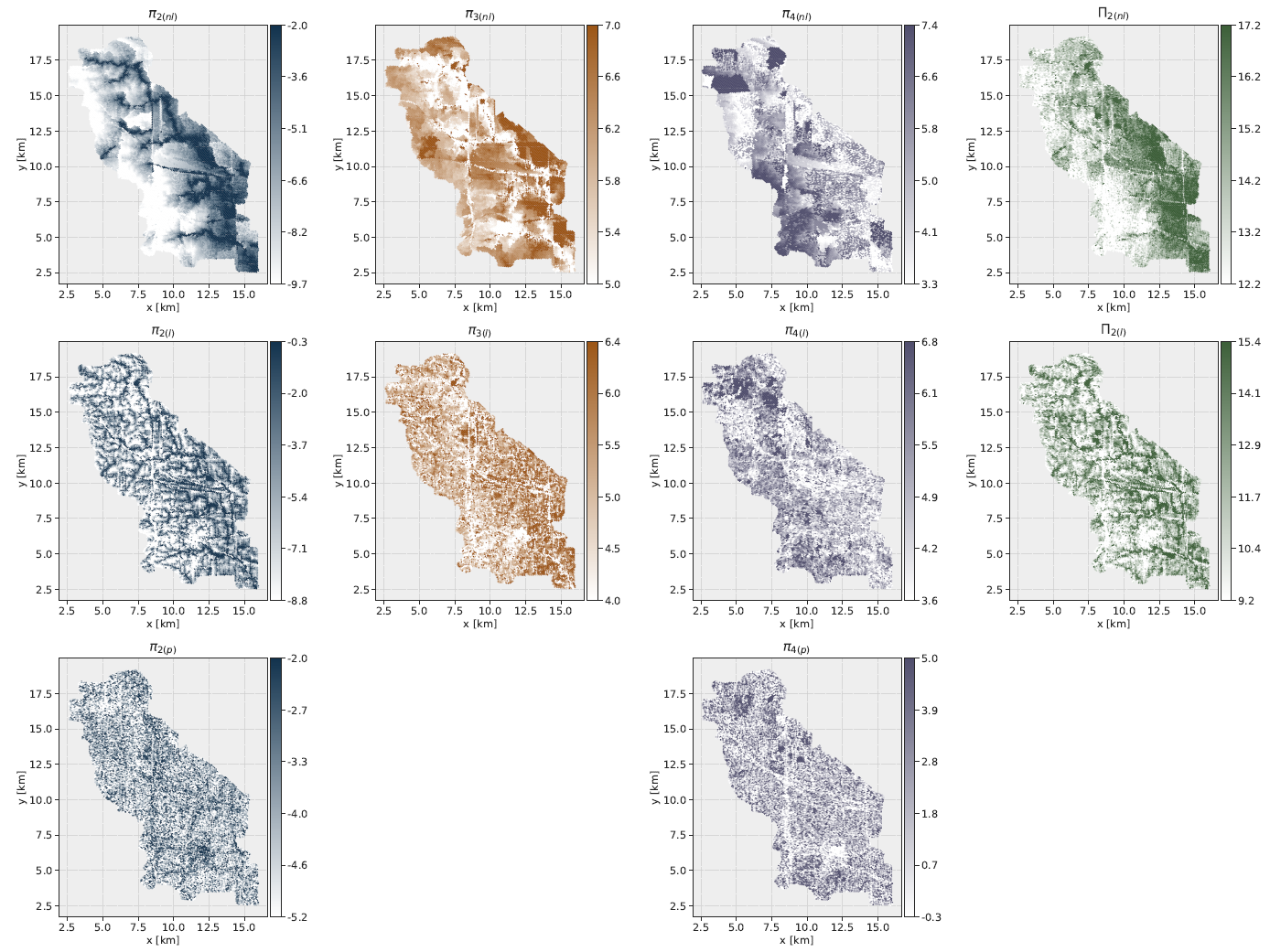}%
\caption{For HUC 071200040403, the dimensionless indices at the different scales of 1) the nonlocal, major streams and rivers with a contributing area greater than 1 sq. mile, i.e., $\pi_{2(nl)}$, $\pi_{3(nl)}$, $\pi_{4(nl)}$, and $\Pi_{2(nl)}$, 2) the localized flow paths (e.g., streets) with a contributing area greater than 0.01 sq. mile, i.e., $\pi_{2(l)}$, $\pi_{3(l)}$, $\pi_{4(l)}$, and $\Pi_{2(l)}$, and 3) at each point, i.e., $\pi_{2(p)}$ and $\pi_{4(p)}$. The  $\pi_2$ terms are based on $Q_{\max}$ derived from the 100-year, 24-hour rainfall event of NOAA Atlas 14 (Table S1), the second quartile, 50th decile temporal distribution of the NOAA Atlas 14, and the extended NRCS-CN method of \citeA{bartlett2015unified1} (see Table S1). \label{fig:7_indices}}
\end{figure}

The dimensionless indices $\pi_2$, $\pi_3$, $\pi_4$, and $\Pi_2$ were calculated at different scales: 1) at each point, i.e. $\pi_{2(p)}$ and $\pi_{4(p)}$, 2) in relation to localized flow paths (e.g. streets), i.e. $\pi_{2(l)}$, $\pi_{3(l)}$, $\pi_{4(l)}$, and $\Pi_{2(l)}$,  and 3) in relation to major streams and rivers that can create non-localized flooding, i.e. $\pi_{2(nl)}$, $\pi_{3(nl)}$, $\pi_{4(nl)}$, and $\Pi_{2(nl)}$. Note that $\pi_3$ and $\Pi_2$ do not exist at a point because the height above drainage is nonexistent without reference to a stream. An example of indices calculated at the 3 different scales is shown on Fig. \ref{fig:7_indices}. 

The dimensionless indices were generally not collinear (Figure S2). Of the 45 pairwise correlations, only 2 were greater than 0.65. The local scale wetness index, $\Pi_{2(l)}$,  was correlated with both the non-local scale wetness index, $\Pi_{2(nl)}$, and the local scale flow capacity ratio, $\pi_{2(l)}$   (0.73 and 0.66, respectively). 
The low correlations of the indices made them suitable for an ML application \cite{murphy2022probabilistic}.

\subsection{Machine Learning Methodology}
\label{sec:ml}

\subsubsection{Pluvial flooding ground-truth (label) data}
In this study, we benchmarked ML training and validation using FEMA's available rainfall/runoff inputs and flood hazard output data from 2D HEC-RAS models, which were run with a spatially uniform but time-varying runoff hyetograph \cite{fema}. These pluvial flood risk data represent a relatively recent addition to FEMA's inventory, which traditionally consists of approximately 1.2 million miles of fluvial stream studies \cite{gao2021flooding}. Since 2019, FEMA has been expanding its flood risk data by incorporating HEC-RAS 2D models, where runoff is applied over a 2D watershed area to enhance flood hazard assessments \cite{tmac2021flooding}.


\subsubsection{Model training and performance}


We hypothesized that using dimensionless input features would improve the generalization of the ML model. To test this hypothesis, watersheds from 2 distinct regions were selected for model training and testing -- 4 watersheds in the Chicago area located in the Des Plaines and Skokie river watersheds (HUC 12 codes starting with 0712) and 4 watersheds in New Jersey in the Hackensack and Passaic river watersheds (with HUC 12 codes starting with 0203) (Fig. S1, Table S1). Compared to the Chicago watersheds, the New Jersey watersheds have a 34-percent decrease in medium- and high-density developed space, an approximately 80-percent increase in topographic slope (on a watershed average basis), and an increase in the density of streams (Table S1). 

Since there was a spatial dependency of the data points, the train-test splits were performed so that entire watersheds were either included or held out from training. For the ML model training, we considered two combinations of these watersheds. Case I investigated whether a machine learning model trained on the watersheds of one region could make meaningful predictions for a completely different region, while Case II evaluated how model performance changed when the training data included watersheds from two different regions.   In Case I, all the Chicago watersheds were used for training, and the model was tested  separately on each of the New Jersey watersheds (50\%-50\% train-test split). In Case II, one watershed from Chicago and one from New Jersey were held out for testing, while the remaining six watersheds were used for training (75\%-25\% train-test split). In each case, a logistic regression model was built for each of the the 10-, 100-, and 1000-year events \cite{cox1958regression}, and each model captured different rainfall ranges: 3.5–6 inches for the 10-year event, 6–8.5 inches for the 100-year event, and 8.5–13.5 inches for the 1000-year event (see Table S2). To benchmark the performance of dimensionless features, logistic regression models also were developed using the dimensional features, which were the derivative data of Fig. \ref{fig:1_process}. Both the dimensional and dimensionless features were standardized, i.e., centered around a mean value and scaled to a unit variance. Correlation matrices for both the dimensionless and dimensional (derivative) data are provided in the supporting information (Figs. S2 and S3).

Logistic regression was selected for the ML model to maintain tractability, interpretability, and to isolate the benefit of the dimensionless features from the selection of the ML model. By using this simpler ML model, we avoided confounding the effects of the dimensionless features with those of a more sophisticated ML model. 
The logistic regression models were trained by optimizing the model weights to minimize the error between the predicted flood classifications and the ground truth data (HEC-RAS-based flood maps). First, the weights were selected to minimize the log-loss with an L1 regularization penalty with a regularization strength of 1 \cite{murphy2022probabilistic}. Secondly, a discrimination threshold used to classify the points as a flooded (class 1) or a not flooded (class 0), and this threshold was estimated by minimizing the difference between the precision and the recall of the predictions. Finally, this optimal discrimination threshold was applied to classify the points within the test watersheds.

Performance was assessed using two metrics: the receiving operator characteristic (ROC) curve and the F$_{\beta}$-score. For the ROC curve, performance was summarized as the area under the curve (AUC) \cite{murphy2022probabilistic}. Generally, an AUC score of 0.8 to 0.9 is considered excellent discrimination, while an AUC score greater than 0.9 is considered outstanding \cite{hosmer2013applied}. An AUC score of 0.5 provides no discrimination between flooded and non-flooded areas and is no better than a random guess. For the F$_{\beta}$-score, the recall was weighted as twice as important as the precision (i.e., $\beta = 2$, see \ref{sec:F1score}). The choice of $\beta=2$ prioritizes the minimization of false negatives. False negatives are particularly costly in flood response as they may direct resources away from flood-prone areas. The F$_2$ score ranges between 0 and 1, with higher values indicating better performance.  The F$_{2}$-score depends on the choice of discrimination threshold and is sensitive to class imbalance and may vary as the event magnitude decreases (i.e., there is less flooded area). Note that we chose the discrimination threshold to balance precision and recall, as described in the previous paragraph.  Differently, the AUC is independent of the discrimination threshold and more robust to class imbalance; however, the AUC can give misleadingly high scores for imbalanced datasets by giving too much weight to the negative class. For such highly imbalanced datasets, the F$_2$ score is better for assessing the detection of the rare positive class (e.g., floods) because the F$_2$ score emphasizes recall. 


\section{Results}
\label{sec:application}

\subsection{Model performance with dimensionless features}
The dimensionless feature ML model performed well according to the AUC scores (Table S3, Fig. S4). Across all events (e.g., 10-, 100-, and 1000-year), the average AUC scores for dimensionless and dimensional features were 0.89 and 0.84 for Case I and 0.89 and 0.87 for Case II, respectively. AUC scores ranged from 0.75 to 0.96. 
The model consistently performed better with dimensionless features compared to with dimensional features, i.e., the derivative data of Fig. \ref{fig:1_process}. This improvement in model performance was most pronounced for the 10-year storm event, with the benefit decreasing for higher return periods as flooding became more widespread. Across storm event return periods, the dimensionless feature model AUC values decreased from 0.9-0.95 (10-yr.) to 0.83-0.90 (100-yr.) and then to 0.83-0.92 (1000-yr.). 

The F$_2$ score is preferred over the AUC for highly imbalanced datasets when detecting the rare positive class (i.e., flooded areas) because it weights the recall more than the precision, i.e., the F$_2$ prioritizes the correct classification of flooded areas. Across all events, the dimensionless feature F$_2$ scores ranged from 0.48 to 0.85, while the dimensional feature F$_2$ score ranged from 0.37 to 0.85. The average F$_2$ scores for the dimensionless and dimensional features were 0.65 and 0.62 for Case I and 0.62 and 0.55 for Case II, respectively (Table \ref{tab:F2}).  Accordingly, the F$_2$ scores indicated better model performance with dimensionless features with an average F$_2$ increase of 9\% in Case I and 15\% in Case II.  Unlike the AUC, the F$_2$ scores indicated better performance for the larger return period events (e.g., 100-year and 1000-year). However, it is important to note that F$_2$ is sensitive to class imbalance, and differences in F$_2$ between events may result from the decreasing class imbalance (i.e., more flooded area) as the event return period increases.

F$_2$ score improvements with the dimensionless features varied across watershed and storm rainfall depth. F$_2$ increased more for the 10-yr. storm than the 100- and 1000-yr. storms on average (20.2\%, 5.3\%, and 7.3\% average increases, respectively). Notably, for watershed 020301030804, the F$_2$ increased by 21.4\% for the 100-year storm and 37.5\% for the 1000-yr. storms. Out of the 18 dimensionless and dimensional F$_2$ score comparisons (Table \ref{tab:F2}), half showed no more than a 5\% difference between the dimensionless and dimensional F$_2$ scores; however, in the other half, the dimensionless features provided an average F$_2$ uplift of 23\%. This average uplift of 23\% showcased how the dimensionless features improved the ML model generalization across watersheds. 

For the 10-year event, all watersheds exhibited an increase in the F$_2$ score, except for watershed 020301030804, where the most notable increase occurred in the 100-year and 1000-year events (Table \ref{tab:F2}). This increase in the F$_2$ across all watersheds suggested no single set of watershed characteristics fully explained this increase---reinforcing the role of dimensionless features in enhancing ML generalization. For watershed 020301030804, the significant uplift in the 100-year and 1000-year events may have resulted from its characteristics. This watershed had a percentage of medium- and high-intensity developed land cover similar to that of the Chicago area training data (see Table S1).  Such urbanization could have led to similar runoff dynamics and channel flows, particularly for the 10-year event. This similarity may have allowed the ML model trained with dimensional data to perform well in this case. Yet, the same watershed also had a much steeper average watershed slope than the Chicago watersheds, which possibly explained the ML model’s underprediction of flooding in the more extreme 100-year and 1000-year events. Overall, this watershed highlighted the role of dimensionless features in enhancing the ML model’s generalization beyond the Chicago area training data.

In Case I where the model was trained in Chicago and tested in New Jersey, the dimensionless features provided an average F$_2$ uplift of 16\% for the 10-year storm, 2.8\% for the 100-year storm, and 4.4\% for the 1000-year storm (Table \ref{tab:F2}). For Case II, in which training data came from both regions, the dimensionless features provided an even greater uplift in the F$_2$ score of 27.7\% for the 10-year storm, 10.1\% for the 100-year storm, and 7\% for the 1000-year storm (Table \ref{tab:F2}). In both casee, the dimensionless features improved ML performance most for the 10-yr. storm and for the watersheds that performed worst with dimensional features. For example, F$_2$ for watershed 020301030804 increased from 0.42 to 0.51 for the 100-yr. storm (and from 0.44 to 0.55 for the 1000-yr. storm). Therefore, dimensionless features provide an advantage compared to dimensional features with respect to generalization, especially for watersheds with poor performance.

\begin{table}[h!]
\begin{minipage}{\textwidth}
\linespread{.5}\selectfont
\caption{F$_2$ scores across 10-yr., 100-yr., and 1000-yr. return period events for Case I and Case II. In Case I, the model was trained on Chicago HUC 12 watersheds and tested on New Jersey HUC 12 watersheds. In Case II, the model was trained on 3 watersheds from each region (6 total) and tested on the remaining 2 watersheds. \label{tab:F2} } 
\noindent
\begin{tabular}{l r r r r r r}
\hline
\noalign{\vskip 0.04in}
 & \multicolumn{2}{c}{10-yr.}  & \multicolumn{2}{c}{100-yr.} & \multicolumn{2}{c}{1000-yr.} \\
\hline
\noalign{\vskip 0.04in}
 \multicolumn{1}{l}{}  & Dim.-less & Dim. & Dim.-less  &  Dim. & Dim.-less & Dim. \\
 \hline
Discrimination Threshold & 0.82 & 0.79 & 0.57 & 0.60 & 0.52 & 0.56 \\
\hline
 Case I F$_2$  &  &   \\
\hline
\noalign{\vskip 0.05in}
              020301030606 &  0.70 & 0.61  & 0.80 & 0.83 & 0.85 & 0.85 \\
              020301030503 &   0.58 & 0.49 & 0.69 & 0.7 & 0.72 & 0.74  \\
              020301030703& 0.61 & 0.45 &  0.57 & 0.6 & 0.61 & 0.64 \\
              020301030804 &  0.66 &  0.68 &  0.51 &  0.42 & 0.55 & 0.44 \\
\hline
\noalign{\vskip 0.04in}
Case II F$_2$ & & & & & &   \\
\hline             
             071200030101 & 0.49 & 0.37 &  0.61 &  0.52 & 0.68 & 0.62 \\
              020301030503  & 0.48 & 0.39 & 0.70  & 0.68  & 0.74 & 0.71   \\
\noalign{\vskip 0.04in}
\hline
\end{tabular}
\\

\end{minipage}
\end{table}

For both Case I and Case II, the optimal discrimination thresholds  (Table \ref{tab:F2}) were relatively similar between the dimensionless and dimensional cases and varied more across storm rainfall depths, as expected.  Compared to dimensional features, the dimensionless features tended to increase the true negative (TN) and decrease the true positive (TP) rates. As an example, for the 100-yr. event, the TN rate increased from 0.82 to 0.90 with dimensionless features (Fig. S5). The TP rate decreased 0.72 to 0.65. Therefore, the dimensionless features improved the ability of the model to correctly predict areas without flooding, while decreasing the ability of the model to correctly predict areas with flooding.


\begin{figure}
\includegraphics[width=6 in]{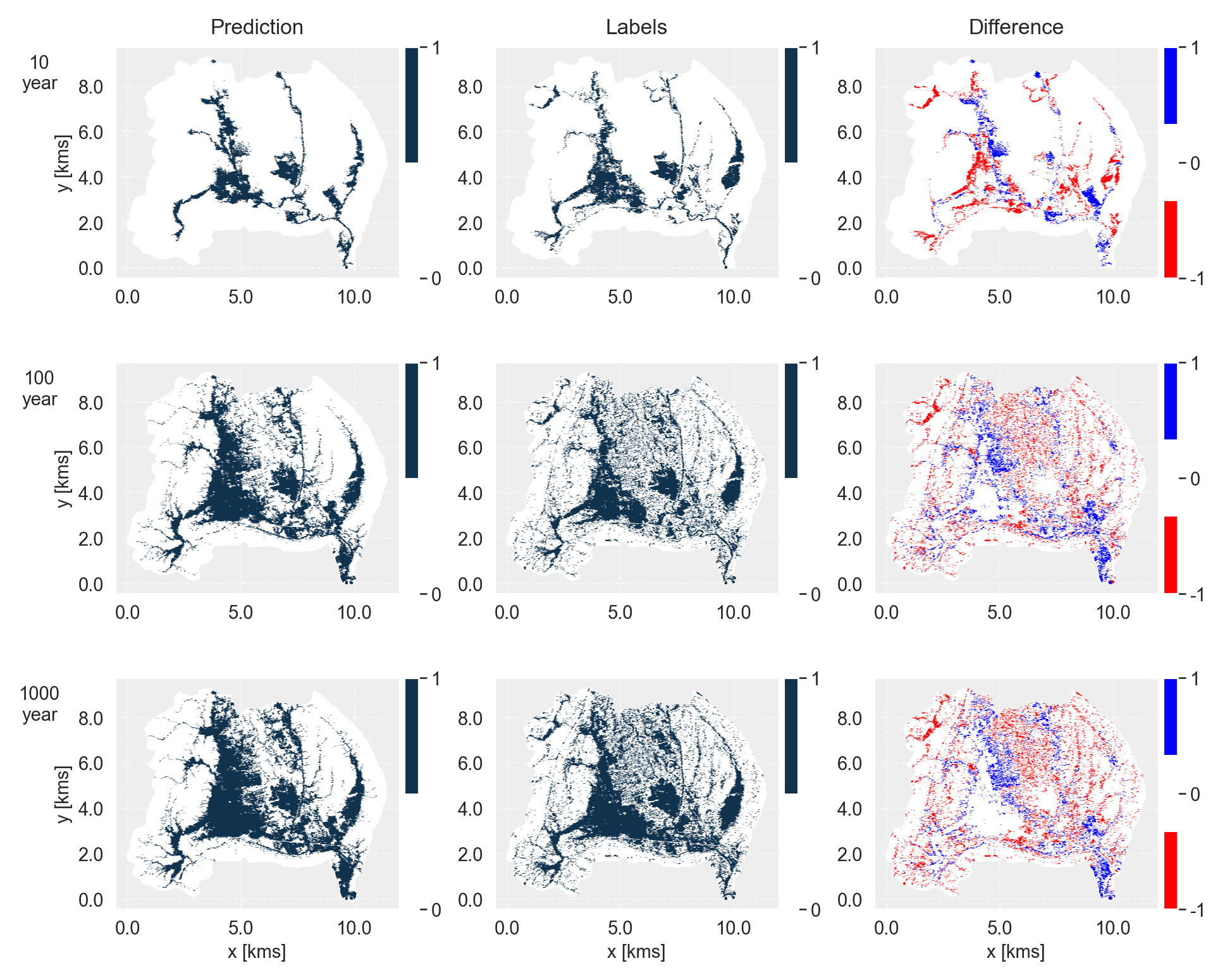}%
\caption{For the Case I test HUC 12 watershed of 020301030503, the ML predictions versus the HEC-RAS generated flood maps (labels) for different return period events, where the discrimination threshold (probability above which a point is flooded) for the 10-year, 1000-year, and 1000-year events were 0.82, 0.57, and 0.53, respectively. The difference was calculated as prediction minus label.\label{fig:9_caseI}}
\end{figure}

\begin{figure}
\includegraphics[width=6 in]{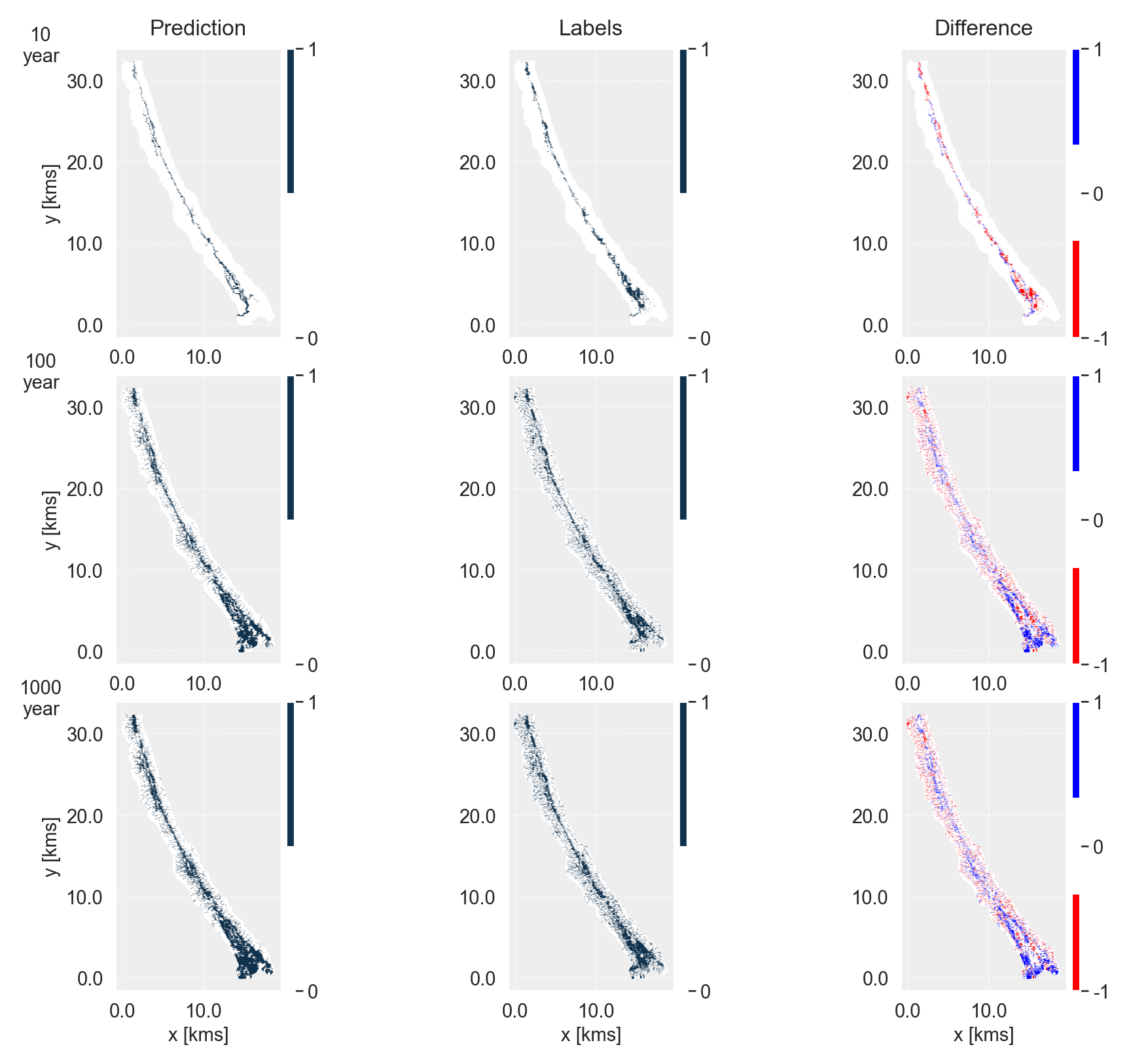}%
\caption{Case II test HUC 12 watershed of 071200030101, the ML predictions versus the HEC-RAS generated flood maps (labels) for different return period events,  where the discrimination threshold (probability above which a point is flooded) for the 10-year, 100-year, and 1000-year events were 0.84, 0.58, and 0.53, respectively. The difference was calculated as prediction minus label.  \label{fig:10_caseII}}
\end{figure}

Figs. \ref{fig:9_caseI} (Case I) and \ref{fig:10_caseII} (Case II) show the ML prediction of flooding compared to the labels derived from 2D HEC-RAS simulations. The ML model captured the spatial pattern of flooding better for the 10-yr. storm compared to the 100- and 1000-yr. storms. The ML model tended to underpredict flood extent (i.e., predict no flooding where the HEC-RAS model predicted flooding). Despite the underpredictions, the ML model generally captured the networked pattern of flash flooding from the HEC-RAS 2D models.



\subsection{Sensitivity to non-local scale threshold}

To study the sensitivity of model performance to the non-local scale threshold, we decreased the contributing area threshold for the non-local flow paths from 1 sq. mile to 0.2 sq. mile and evaluated the absolute change in the AUC (Table \ref{tab:cases_change}). Generally, performance decreased, with the most significant performance decreases occurring for the 10-year event (Table \ref{tab:cases_change}). The 10-year event had the most dramatic performance decrease because most of the flooding for a 10-year event is centered around the major rivers and streams, and a lower non-local threshold (of 0.2 sq. miles) does not define major rivers and streams well. The magnitude of the change in some cases demonstrated the importance of optimizing the thresholds for flow path delineation in relation to the performance of the ML model. 

\begin{table}[h!]\centering
\begin{minipage}{\textwidth}
\linespread{.5}\selectfont
\caption{Sensitivity of AUC to the non-local scale threshold\label{tab:cases_change} }
\noindent
\begin{tabular}{l c c c }
\hline
\noalign{\vskip 0.04in}
\noalign{\vskip 0.04in}
  \multicolumn{1}{l|}{}  &  \multicolumn{3}{c}{Test AUC change} \\
\multicolumn{1}{l|}{Case I}    & 10-yr. & 100-yr. & 1000-yr.    \\
\hline
\noalign{\vskip 0.05in}
\multicolumn{1}{l|}{020301030606}  & -0.01 & -0.01  & -0.01    \\
\multicolumn{1}{l|}{020301030503}  & -0.05 & -0.02  & -0.02 \\
\multicolumn{1}{l|}{020301030703}  & -0.08 & -0.01 & -0.01 \\
\multicolumn{1}{l|}{020301030804}  & -0.08  & 0 & 0 \\
\hline
\noalign{\vskip 0.04in}
\multicolumn{4}{l}{Case II} \\
\hline             
\multicolumn{1}{l|}{071200030101}  & -0.14 & -0.05   & -0.05 \\
\multicolumn{1}{l|}{020301030503}  & -0.05   & -0.02 & -0.03 \\ 
\noalign{\vskip 0.04in}
\hline
\end{tabular}
\\
\emph{Note}: Absolute change in AUC after reducing the non-local contributing area threshold from 1 mi$^2$ to 0.2 mi$^2$. \\
\end{minipage}
\end{table}

\section{Discussion}

Recent studies have demonstrated that dimensionless features can improve ML generalization \cite{gunaratnam2003improving,oppenheimer2023multi} and computational efficiency \cite{hu2019rapid}. Dimensionless features improve generalization by redefining the training space such that a larger number of test cases fall within the training set \cite{oppenheimer2023multi}. This reduces extrapolation for new cases. In a similar vein, Hu et al. (2019) used proper orthogonal decomposition (POD) and singular value decomposition (SVD) to define a reduced-order space in which a long short-term memory (LSTM) model was trained to predict flood depth. They noted improved computational efficiency as the LSTM was trained with a smaller number of features. In contrast, we used the Buckingham $\Pi$ theorem to define dimensionless input features in a reduced-order space and showed that dimensionless features perform better than traditional dimensional features with respect to model generalization between regions. Our model was also more computationally efficient as the numerous dimensional derivative data input layers of Figure \ref{fig:1_process} are combined into 4 dimensionless features prior to model training (for each of the point, local, and non-local features). While previous studies have applied Buckingham $\Pi$ theorem to hydrology \cite<e.g.,>{porporato2022hydrology}, this is the first study, to our knowledge, to apply Buckingham $\Pi$ theorem for ML in flood mapping, and we anticipate wide applicability in flood and streamflow forecasting. 

The dimensionless features performed comparable to previous ML flood modeling efforts (e.g., \citeA{manfreda2014investigation,samela2016based}), even when the model was trained and tested in different regions (e.g., Chicago and New Jersey). Therefore, the use of dimensionless features in ML flood modeling has potential to improve model generalization across terrains. Other approaches have been proposed to address model generalization, such as including multi-scale contextual terrain information \cite{cache2024enhancing}, reduced-order feature selection \cite{hu2019rapid,pakdehi2023transferability}, transfer learning \cite{seleem2022transferability}, and the application of convolutional neural networks (CNNs) \cite{guo2022data}. Here, we used logistic regression to isolate and maintain focus on performance of the dimensionless features. However, we anticipate that multi-layer neural networks would improve model performance.


The dimensionless indices when coupled to the ML model allow for a rapid prediction of pluvial (flash flood) extents. The dimensionless features are largely based on static terrain data (e.g., DEM, land cover, stream network) and, therefore, only need to be processed once. The only non-static inputs are the antecedent moisture condition (represented by a CN value) and storm precipitation. Based on the CN, the GIUH is quickly recalculated (in seconds) and convolved with the precipitation (also in seconds) to recalculate the flows for the $\pi_2$ terms. In future work, non-static rainfall and evapotranspiration climatology could be incorporated to account for seasonality and climate change in the $\Pi_2$ wetness index term. The processing of the dimensionless data takes about 2-3 hours for DEM resolutions of 1- to 3-meters. Once the derivative data are created, they can be combined with runoff hyetographs in seconds and new ML results produced within minutes. Thus, runoff hyetographs from the latest weather forecasts (e.g., from the High-Resolution Rapid Refresh (HRRR) model \cite{dowell2022high}) could be rapidly transformed into maps that inform flash flood warnings. Such rapid, high-resolution mapping would be an improvement over current flash flood warnings that generally are given on a regional basis without specificity. This lack of specificity makes it difficult to truly understand the best actions to preserve life and maintain safety during a flash flood warning.



The definition of the dimensionless $\pi$ and $\Pi$ terms is not unique --- multiple valid formulations exist under Buckingham $\Pi$ theorem.  Future work could explore and quantify the performance of different formulations of the dimensionless indices. Identifying further performance improvements could involve 1) exploring optimal thresholds for delineating flow paths at different scales to better capture flood processes \cite{sangireddy2016geonet} and 2) resolving reach-averaged hydraulics at a subwatershed level within each HUC 12 watershed. Currently, reach-average hydraulics are calculated across an entire HUC 12 watershed. Additionally, advanced image processing techniques could be integrated into the model, either as a pre-processing step or as part of the machine learning process itself. A fully convolutional neural network (e.g., the U-Net ML model for image segmentation \cite{ronneberger2015u}) could further improve the flash flood prediction by capturing complex spatial patterns, offering a more flexible alternative to the current logistic regression framework.

In this study for the $\pi_2$ terms, the flow hydrograph was estimated using the geomorphological instantaneous unit hydrograph (GIUH). By convolving the GIUH for the runoff-producing area with a rainfall hyetograph, we generated a watershed-wide flow signature that reflects the relationship between flow and the contributing area (e.g., Fig. \ref{fig:5_flow_vs_area}). This approach is advantageous because it relies on data that are readily available at both continental and global scales, such as DEM and land cover, making it scalable for broad applications. Furthermore, the performance of the ML model developed in this work, which incorporates the GIUH, demonstrates the potential of the GIUH to improve regional streamflow estimation \cite{capesius2009regional}. Thus, integrating GIUH attributes into regional streamflow regression models could significantly enhance performance compared to the traditional regression-based methods. 

Lastly, a number of assumptions were employed here that warrant further study. First, a small set of training and testing watersheds was used here to test the hypothesis that the dimensionless features improved model generalization. Secondly, the model was trained and tested on HEC-RAS model output, not observed flood events. Lastly, models were generalized only across regions and not across storm depths. Future testing and development of dimensionless features for flood modeling should expand to a larger number of watershed contexts, test against observed flood extents, and address generalization across events. In this study, a redefinition of the training space was observed in the model's performance, particularly with the 10-year event, where the predictive performance dropped when the model was trained across two regions, as compared to when it was trained within a single region (Chicago) and applied to a different region (New Jersey). Future work should examine the performance of the ML model in various hydroclimates by training it in one hydroclimate and testing it in a significantly different one. Additionally, further analysis should assess how model accuracy is influenced by the inclusion or exclusion of different scales in hydrographically derived variables.

\section{Concluding Remarks}
Dimensionless features can improve ML model performance, especially when generalizing to never-before-seen conditions. By reformulating a physical problem into a dimensionless context with Buckingham $\Pi$ theorem, we may capture the similarity of a process across different environments,  conditions, and scales. With respect to flood mapping with ML, this approach can be expanded to other regions and catchments, a wider range of storm events, and other flooding mechanisms such as riverine, coastal, and compound flooding. 
We also anticipate that this approach will find general utility in assessing different geohazard risks.

Future research could 1) refine the proposed framework by exploring alternative hydrological formulations beyond the quasi-steady-state assumptions of TOPMODEL, 2) optimize flow accumulation thresholds for stream delineation across scales, and 3) expand the dimensionless $\Pi$ terms to account for engineered infrastructure such as levees and dams. In addition, a systematic statistical characterization of watershed features could aid in identifying new watersheds with features that are in sample with the training dataset---potentially providing a  path forward for estimating the ML model performance in new never-before-seen locations. Further improvements could integrate satellite-derived flood extents into the ML training, enhancing model generalization \cite{yang2021high}. By bridging hydrological modeling, remote sensing, and data-driven approaches, these advancements would strengthen the physical interpretability, transferability, and scalability of the ML approach presented in this study, improving its applicability across diverse hydrological and climatic contexts.

The concepts described here can be applied to the broader area of reproducibility in ML models, which includes feature engineering, generalizability, interpretable ML, and efficiency and scalability.  The physical processes described by the input features often involve relationships between variables that are maintained across different scales and conditions (e.g., climate, fluid dynamics, financial systems). Derivation of these relationships from fundamental physical principles (via the Buckingham $\Pi$ theorem) is an innovative method to improve model transferability. Furthermore, this approach can be considered more interpretable, as the input features are meaningfully constrained by the underlying physics, represent known controls on flooding extent, and are dimensionally independent. Lastly, models with a minimal number of features that generalize better will improve efficiency and scalability. Models with reduced computational demand while maintaining accuracy and generalization can be deployed in real-time scenarios. The methods described in the article not only advance flood inundation mapping but also offer insights into general ML practices. The proposed framework provides a path forward for developing more robust, scalable, and reproducible ML models that can be applied to diverse and dynamic environments.

\appendix




\section{Hydraulic Radius Adjustment factor}
\label{sec:App_Rh}
The hydraulic radius is the cross sectional area divided by the wetted perimeter. Accordingly, the factor adjusting the hydraulic radius, $\alpha_{R_h}$, is the factor for adjusting the cross sectional area, $\alpha_A = \frac{2y_d}{\overline{w}}$, multiplied by a factor for adjusting the wetted perimeter. Note that  $\alpha_A$ is the ratio of river width (estimated as $2\cdot y_d$) to the reach average width, $\overline{w}$. Here, the wetted perimeter factor is a weighted combination of the factors for a triangular cross section, $\alpha_T$, and a parabolic cross section, $\alpha_P$, i.e.,

\begin{align}
\label{eq:alpha_2}
    \alpha_{R_h} = 
    &\left\{
    \begin{array}{lll}
         \alpha_{T} \alpha_A & &  A < A_T \\
         \alpha_{T} \alpha_A\frac{A-A_T}{A_P-A_T}+ \alpha_{P} \alpha_A \left(1 - \frac{A-A_T}{A_P-A_T}\right) \quad\quad  & A_T \leq   & A < A_P \\
         \alpha_{P} \alpha_A  &   A_P \leq & A,
    \end{array}    
    \right.\nonumber
\end{align}
where the overall factor varies linearly depending on how the cross sectional area, $A$, compares to an assumed triangular cross section,  $A_T = y_d\cdot z_d$, and parabolic cross section, $A_P = \frac{2}{3}2 y_d\cdot z_d$--both calculated based on the distance from drainage, $y_d$, and height above drainage, $z_d$.  The wetted perimeter factors for the triangular cross section, $\alpha_T$, and parabolic cross section, $\alpha_P$, respectively are \cite{chow1959open},

\begin{align}
    \alpha_T &= \frac{2 z_d\sqrt{\frac{\overline{w}}{z_d}+1}}{2 z_d\sqrt{\frac{2 y_{d}}{z_d}+1}}, \\
    \alpha_P & = \frac{\overline{w} +\frac{8 z_d^2}{3 \overline{w}}}{2 y_d +\frac{8 z_d^2}{3\cdot 2y_d}},
\end{align}
where each factor is the wetted perimeter based on the reach averaged width, $\overline{w}$, divided by the wetted perimeter based on the channel width (estimated as $2\cdot y_d$).

\section{F$_{\beta}$-score}
\label{sec:F1score}

Binary classification produces predictions of true positives (TP), true negatives (TN), false positives (FP), and false negatives (FN). A successful statistical inference model maximizes TP and TN relative to FP and FN. These four categories (i.e., TP, TN, FP, and FN) are preferred over raw accuracy for imbalanced datasets such as flood datasets where, in most cases, dry areas exceed flooded areas in extent. The raw accuracy is calculated as

\begin{equation}
    \text{accuracy} = \frac{\text{TP} + \text{TN}}{\text{All data points}} = \frac{\text{TP} + \text{TN}}{\text{TP} + \text{TN} + \text{FP} + \text{FN}}.
    \label{eq:raw_accuracy}
\end{equation}
Assessing model performance on the raw accuracy of Equation \ref{eq:raw_accuracy} is inappropriate for imbalanced data because it allows for models to perform poorly on the minority class. For example, if a geospatial dataset is 5\% flooded by extent, one can train a model with 95\% accuracy by predicting all pixels as dry.

The F$_{\beta}$-score is an improvement on raw accuracy for imbalanced data. Once the model is trained, it is evaluated by calculating the F$_{\beta}$-score on the test set, which weights recall as $\beta>0$ more significant than precision, i.e., \cite{murphy2022probabilistic}

\begin{align}
F_{\beta}=\frac{1}{\frac{1}{1+\beta^2}\frac{1}{P}+\frac{\beta^2}{1+\beta^2}\frac{1}{R}},
\end{align}
where the precision, $P$, is written as

\begin{equation}
    P = \frac{\text{TP}}{\text{TP} + \text{FP}},
    \label{eq:precision}
\end{equation}
and the recall, $R$, is written as

\begin{equation}
    R = \frac{\text{TP}}{\text{TP} + \text{FN}}.
    \label{eq:recall}
\end{equation}
When the precision and recall are equally weighted, i.e., $\beta=1$, then we retrieve the F$_1$-score that is the harmonic mean of the precision, $P$, and recall, $R$, i.e.,

\begin{equation}
    F_1 = \frac{2\, P\, R}{P+R}.
    \label{eq:f1_score}
\end{equation}

When the recall is weighted twice as important as the precision, i.e., $\beta=2$, then we retrieve the F$_2$-score,

\begin{equation}
    F_2 = \frac{P\, R}{0.8 P+0.2 R}.
    \label{eq:f2_score}
\end{equation}

Equations \ref{eq:precision} and \ref{eq:recall} both have TP in their numerators and denominators which suggests that the F$_{\beta}$ score prioritizes the positive class over the negative one. This is appropriate when the imbalance in the data is in the negative class' favor; this is when the positive class is in the minority. More intuitively, the recall, $R$, also may be thought of as a 'hit rate', i.e., the fraction of predictions that capture flooding accurately in comparison to the assumed ground truth data, while 1 minus the precision, $1-P$, may be though of as a 'false alarm rate', i.e., the fraction of predictions that falsely predicted flooding in comparison to the assumed ground truth data.

\begin{acknowledgments}
Funding for this work was provided by a cooperative endeavor agreement between The Water Institute and the Louisiana Office of Community Development as part of the Louisiana Watershed Initiative (LWI) funded via Catalog of Federal Domestic Assistance (CFDA) 14.228 Grant B-18-DP-22-0001, as well as by Stantec and The Water Research Foundation (WRF) project 5084, \emph{Holistic and Innovative Approaches for Flood Mitigation Planning and Modeling under Extreme Wet Weather Events and Climate Impacts}. We thank Harry Zhang of WRF for his support and encouragement. We thank Marco Maneta, Salvatore Manfreda, and another anonymous reviewer  for their insightful and constructive comments.

\emph{Data Availability Statement}. The data supporting this manuscript is available at the locations shown in Table \ref{tab:data}. The DEM data is available from the USGS 3DEP elevation product \cite{USGS_3DEP_DEM}; hydrography data, including watershed boundaries, are available from the USGS NHD \cite{USGS_NHD}; the land use, roads, and impervious descriptor data are available from the NLCD data products \cite{USGS_2024_NLCD}; soil data are available from the SSURGO database \cite{SSURGO_Database}; flood maps and data are available from the FEMA map service center \cite{fema}; rainfall extremes are available from NOAA Atlas 14 \cite{bonnin2004atlas}; and climatology data are available from Daymet \cite{thornton2022daymet}. The data processing was performed with GRASS GIS \cite{GRASS_GIS_software}, while the machine learning was performed using Scikit-learn \cite{scikit-learn}. The data and the script used to analyze the data and create the figures are available on Zenodo \cite{bartlett_physically-based_2025}.

\end{acknowledgments}

\bibliography{References}

\begin{thebibliography}{}

\bibitem [\protect \citeauthoryear {%
Ashley%
\ \BBA {} Ashley%
}{%
Ashley%
\ \BBA {} Ashley%
}{%
{\protect \APACyear {2008}}%
}]{%
ashley2008flood}
\APACinsertmetastar {%
ashley2008flood}%
\begin{APACrefauthors}%
Ashley, S\BPBI T.%
\BCBT {}\ \BBA {} Ashley, W\BPBI S.%
\end{APACrefauthors}%
\unskip\
\newblock
\APACrefYearMonthDay{2008}{}{}.
\newblock
{\BBOQ}\APACrefatitle {Flood fatalities in the United States} {Flood fatalities
  in the united states}.{\BBCQ}
\newblock
\APACjournalVolNumPages{Journal of Applied Meteorology and
  Climatology}{47}{3}{805--818}.
\PrintBackRefs{\CurrentBib}

\bibitem [\protect \citeauthoryear {%
Bartlett%
\ \protect \BOthers {.}}{%
Bartlett%
\ \protect \BOthers {.}}{%
{\protect \APACyear {2025}}%
}]{%
bartlett_physically-based_2025}
\APACinsertmetastar {%
bartlett_physically-based_2025}%
\begin{APACrefauthors}%
Bartlett, M\BPBI S.%
, Blitterswyk, J\BPBI V.%
, Farella, M.%
, Li, J.%
, Parolari, A.%
, Krishnamoorthy, L.%
\BCBL {}\ \BBA {} Mrad, A.%
\end{APACrefauthors}%
\unskip\
\newblock
\APACrefYearMonthDay{2025}{{\APACmonth{03}}}{}.
\newblock
\APACrefbtitle {Physically-based dimensionless features for pluvial flood
  mapping with machine learning: Dataset and computational notebook [{Dataset,
  Computational Notebook}].} {Physically-based dimensionless features for
  pluvial flood mapping with machine learning: Dataset and computational
  notebook [{Dataset, Computational Notebook}].}
\newblock
\APACaddressPublisher{}{Zenodo}.
\newblock
\begin{APACrefURL} \url{https://zenodo.org/records/15045598} \end{APACrefURL}
\newblock
\begin{APACrefDOI} \doi{10.5281/zenodo.15045598} \end{APACrefDOI}
\PrintBackRefs{\CurrentBib}

\bibitem [\protect \citeauthoryear {%
Bartlett%
, Parolari%
, McDonnell%
\BCBL {}\ \BBA {} Porporato%
}{%
Bartlett%
\ \protect \BOthers {.}}{%
{\protect \APACyear {2017}}%
}]{%
bartlett2017reply}
\APACinsertmetastar {%
bartlett2017reply}%
\begin{APACrefauthors}%
Bartlett, M\BPBI S.%
, Parolari, A\BPBI J.%
, McDonnell, J.%
\BCBL {}\ \BBA {} Porporato, A.%
\end{APACrefauthors}%
\unskip\
\newblock
\APACrefYearMonthDay{2017}{}{}.
\newblock
{\BBOQ}\APACrefatitle {Reply to comment by {F}red {L}. {O}gden et al. on
  ‘‘{B}eyond the {SCS-CN} method: {A} theoretical framework for spatially
  lumped rainfall-runoff response,’’} {Reply to comment by {F}red {L}.
  {O}gden et al. on ‘‘{B}eyond the {SCS-CN} method: {A} theoretical
  framework for spatially lumped rainfall-runoff response,’’}.{\BBCQ}
\newblock
\APACjournalVolNumPages{Water Resources Research}{53}{7}{6351-6354}.
\PrintBackRefs{\CurrentBib}

\bibitem [\protect \citeauthoryear {%
Bartlett%
, Parolari%
, McDonnell%
\BCBL {}\ \BBA {} Porporato%
}{%
Bartlett%
\ \protect \BOthers {.}}{%
{\protect \APACyear {2016}}%
{\protect \APACexlab {{\protect \BCnt {1}}}}}]{%
bartlett2015unified1}
\APACinsertmetastar {%
bartlett2015unified1}%
\begin{APACrefauthors}%
Bartlett, M\BPBI S.%
, Parolari, A\BPBI J.%
, McDonnell, J\BPBI J.%
\BCBL {}\ \BBA {} Porporato, A.%
\end{APACrefauthors}%
\unskip\
\newblock
\APACrefYearMonthDay{2016{\protect \BCnt {1}}}{}{}.
\newblock
{\BBOQ}\APACrefatitle {Beyond the {SCS-CN} method: {A} theoretical framework
  for spatially lumped rainfall-runoff response} {Beyond the {SCS-CN} method:
  {A} theoretical framework for spatially lumped rainfall-runoff
  response}.{\BBCQ}
\newblock
\APACjournalVolNumPages{Water Resources Research}{52}{6}{4608--4627}.
\PrintBackRefs{\CurrentBib}

\bibitem [\protect \citeauthoryear {%
Bartlett%
, Parolari%
, McDonnell%
\BCBL {}\ \BBA {} Porporato%
}{%
Bartlett%
\ \protect \BOthers {.}}{%
{\protect \APACyear {2016}}%
{\protect \APACexlab {{\protect \BCnt {2}}}}}]{%
bartlett2015unified2}
\APACinsertmetastar {%
bartlett2015unified2}%
\begin{APACrefauthors}%
Bartlett, M\BPBI S.%
, Parolari, A\BPBI J.%
, McDonnell, J\BPBI J.%
\BCBL {}\ \BBA {} Porporato, A.%
\end{APACrefauthors}%
\unskip\
\newblock
\APACrefYearMonthDay{2016{\protect \BCnt {2}}}{}{}.
\newblock
{\BBOQ}\APACrefatitle {Framework for event-based semidistributed modeling that
  unifies the {SCS}-{CN} method, {VIC}, {PDM}, and {TOPMODEL}} {Framework for
  event-based semidistributed modeling that unifies the {SCS}-{CN} method,
  {VIC}, {PDM}, and {TOPMODEL}}.{\BBCQ}
\newblock
\APACjournalVolNumPages{Water Resources Research}{52}{9}{7036--7052}.
\PrintBackRefs{\CurrentBib}

\bibitem [\protect \citeauthoryear {%
Bentivoglio%
, Isufi%
, Jonkman%
\BCBL {}\ \BBA {} Taormina%
}{%
Bentivoglio%
\ \protect \BOthers {.}}{%
{\protect \APACyear {2022}}%
}]{%
bentivoglio2022deep}
\APACinsertmetastar {%
bentivoglio2022deep}%
\begin{APACrefauthors}%
Bentivoglio, R.%
, Isufi, E.%
, Jonkman, S\BPBI N.%
\BCBL {}\ \BBA {} Taormina, R.%
\end{APACrefauthors}%
\unskip\
\newblock
\APACrefYearMonthDay{2022}{}{}.
\newblock
{\BBOQ}\APACrefatitle {Deep learning methods for flood mapping: a review of
  existing applications and future research directions} {Deep learning methods
  for flood mapping: a review of existing applications and future research
  directions}.{\BBCQ}
\newblock
\APACjournalVolNumPages{Hydrology and Earth System Sciences
  Discussions}{2022}{}{1--50}.
\PrintBackRefs{\CurrentBib}

\bibitem [\protect \citeauthoryear {%
Beven%
}{%
Beven%
}{%
{\protect \APACyear {2006}}%
}]{%
beven2006manifesto}
\APACinsertmetastar {%
beven2006manifesto}%
\begin{APACrefauthors}%
Beven, K.%
\end{APACrefauthors}%
\unskip\
\newblock
\APACrefYearMonthDay{2006}{}{}.
\newblock
{\BBOQ}\APACrefatitle {A manifesto for the equifinality thesis} {A manifesto
  for the equifinality thesis}.{\BBCQ}
\newblock
\APACjournalVolNumPages{Journal of hydrology}{320}{1}{18--36}.
\PrintBackRefs{\CurrentBib}

\bibitem [\protect \citeauthoryear {%
Beven%
}{%
Beven%
}{%
{\protect \APACyear {2012}}%
}]{%
beven2012rainfall}
\APACinsertmetastar {%
beven2012rainfall}%
\begin{APACrefauthors}%
Beven, K.%
\end{APACrefauthors}%
\unskip\
\newblock
\APACrefYear{2012}.
\newblock
\APACrefbtitle {Rainfall-Runoff Modelling: The Primer} {Rainfall-runoff
  modelling: The primer}.
\newblock
\APACaddressPublisher{}{Wiley}.
\PrintBackRefs{\CurrentBib}

\bibitem [\protect \citeauthoryear {%
Beven%
\ \BBA {} Freer%
}{%
Beven%
\ \BBA {} Freer%
}{%
{\protect \APACyear {2001}}%
}]{%
beven2001equifinality}
\APACinsertmetastar {%
beven2001equifinality}%
\begin{APACrefauthors}%
Beven, K.%
\BCBT {}\ \BBA {} Freer, J.%
\end{APACrefauthors}%
\unskip\
\newblock
\APACrefYearMonthDay{2001}{}{}.
\newblock
{\BBOQ}\APACrefatitle {Equifinality, data assimilation, and uncertainty
  estimation in mechanistic modelling of complex environmental systems using
  the GLUE methodology} {Equifinality, data assimilation, and uncertainty
  estimation in mechanistic modelling of complex environmental systems using
  the glue methodology}.{\BBCQ}
\newblock
\APACjournalVolNumPages{Journal of hydrology}{249}{1}{11--29}.
\PrintBackRefs{\CurrentBib}

\bibitem [\protect \citeauthoryear {%
Beven%
\ \BBA {} Kirkby%
}{%
Beven%
\ \BBA {} Kirkby%
}{%
{\protect \APACyear {1979}}%
}]{%
beven1979physically}
\APACinsertmetastar {%
beven1979physically}%
\begin{APACrefauthors}%
Beven, K.%
\BCBT {}\ \BBA {} Kirkby, M.%
\end{APACrefauthors}%
\unskip\
\newblock
\APACrefYearMonthDay{1979}{}{}.
\newblock
{\BBOQ}\APACrefatitle {A physically based, variable contributing area model of
  basin hydrology/Un mod{\`e}le {\`a} base physique de zone d'appel variable de
  l'hydrologie du bassin versant} {A physically based, variable contributing
  area model of basin hydrology/un mod{\`e}le {\`a} base physique de zone
  d'appel variable de l'hydrologie du bassin versant}.{\BBCQ}
\newblock
\APACjournalVolNumPages{Hydrological Sciences Journal}{24}{1}{43--69}.
\PrintBackRefs{\CurrentBib}

\bibitem [\protect \citeauthoryear {%
Bonnin%
\ \protect \BOthers {.}}{%
Bonnin%
\ \protect \BOthers {.}}{%
{\protect \APACyear {2004}}%
}]{%
bonnin2004atlas}
\APACinsertmetastar {%
bonnin2004atlas}%
\begin{APACrefauthors}%
Bonnin, G\BPBI M.%
, Martin, D.%
, Lin, B.%
, Parzybok, T.%
, Yekta, M.%
\BCBL {}\ \BBA {} Riley, D.%
\end{APACrefauthors}%
\unskip\
\newblock
\APACrefYearMonthDay{2004}{}{}.
\newblock
\APACrefbtitle {Atlas 14 {P}recipitation-{F}requency {A}tlas of the {U}nited
  {S}tates} {Atlas 14 {P}recipitation-{F}requency {A}tlas of the {U}nited
  {S}tates}\ (\BVOL~2).
\newblock
\APACaddressPublisher{Silver Spring, Maryland}{}.
\newblock
\begin{APACrefURL} \url{https://hdsc.nws.noaa.gov/pfds/} \end{APACrefURL}
\newblock
\APACrefnote{Revised 2004}
\PrintBackRefs{\CurrentBib}

\bibitem [\protect \citeauthoryear {%
Brunner%
}{%
Brunner%
}{%
{\protect \APACyear {2016}}%
}]{%
brunner2016hec}
\APACinsertmetastar {%
brunner2016hec}%
\begin{APACrefauthors}%
Brunner, G\BPBI W.%
\end{APACrefauthors}%
\unskip\
\newblock
\APACrefYearMonthDay{2016}{}{}.
\newblock
{\BBOQ}\APACrefatitle {{HEC-RAS} {R}iver {A}nalysis {S}ystem. {H}ydraulic
  {R}eference {M}anual. Version 5.0} {{HEC-RAS} {R}iver {A}nalysis {S}ystem.
  {H}ydraulic {R}eference {M}anual. version 5.0}{\BBCQ}\
  [\bibcomputersoftwaremanual].
\PrintBackRefs{\CurrentBib}

\bibitem [\protect \citeauthoryear {%
Bui%
\ \protect \BOthers {.}}{%
Bui%
\ \protect \BOthers {.}}{%
{\protect \APACyear {2018}}%
}]{%
bui2018novel}
\APACinsertmetastar {%
bui2018novel}%
\begin{APACrefauthors}%
Bui, D\BPBI T.%
, Panahi, M.%
, Shahabi, H.%
, Singh, V\BPBI P.%
, Shirzadi, A.%
, Chapi, K.%
\BDBL {}others%
\end{APACrefauthors}%
\unskip\
\newblock
\APACrefYearMonthDay{2018}{}{}.
\newblock
{\BBOQ}\APACrefatitle {Novel hybrid evolutionary algorithms for spatial
  prediction of floods} {Novel hybrid evolutionary algorithms for spatial
  prediction of floods}.{\BBCQ}
\newblock
\APACjournalVolNumPages{Scientific reports}{8}{1}{1--14}.
\PrintBackRefs{\CurrentBib}

\bibitem [\protect \citeauthoryear {%
Cache%
\ \protect \BOthers {.}}{%
Cache%
\ \protect \BOthers {.}}{%
{\protect \APACyear {2024}}%
}]{%
cache2024enhancing}
\APACinsertmetastar {%
cache2024enhancing}%
\begin{APACrefauthors}%
Cache, T.%
, Gomez, M\BPBI S.%
, Beucler, T.%
, Blagojevic, J.%
, Leitao, J\BPBI P.%
\BCBL {}\ \BBA {} Peleg, N.%
\end{APACrefauthors}%
\unskip\
\newblock
\APACrefYearMonthDay{2024}{}{}.
\newblock
{\BBOQ}\APACrefatitle {Enhancing generalizability of data-driven urban flood
  models by incorporating contextual information} {Enhancing generalizability
  of data-driven urban flood models by incorporating contextual
  information}.{\BBCQ}
\newblock
\APACjournalVolNumPages{Hydrology and Earth System Sciences
  Discussions}{2024}{}{1--23}.
\PrintBackRefs{\CurrentBib}

\bibitem [\protect \citeauthoryear {%
Capesius%
\ \BBA {} Stephens%
}{%
Capesius%
\ \BBA {} Stephens%
}{%
{\protect \APACyear {2009}}%
}]{%
capesius2009regional}
\APACinsertmetastar {%
capesius2009regional}%
\begin{APACrefauthors}%
Capesius, J\BPBI P.%
\BCBT {}\ \BBA {} Stephens, V\BPBI C.%
\end{APACrefauthors}%
\unskip\
\newblock
\APACrefYear{2009}.
\newblock
\APACrefbtitle {Regional regression equations for estimation of natural
  streamflow statistics in Colorado} {Regional regression equations for
  estimation of natural streamflow statistics in colorado}.
\newblock
\APACaddressPublisher{}{US Department of the Interior, US Geological Survey}.
\PrintBackRefs{\CurrentBib}

\bibitem [\protect \citeauthoryear {%
Chow%
}{%
Chow%
}{%
{\protect \APACyear {1959}}%
}]{%
chow1959open}
\APACinsertmetastar {%
chow1959open}%
\begin{APACrefauthors}%
Chow, V\BPBI T.%
\end{APACrefauthors}%
\unskip\
\newblock
\APACrefYearMonthDay{1959}{}{}.
\newblock
{\BBOQ}\APACrefatitle {Open-channel hydraulics} {Open-channel
  hydraulics}.{\BBCQ}
\newblock
\APACjournalVolNumPages{McGraw-Hill civil engineering series}{}{}{}.
\PrintBackRefs{\CurrentBib}

\bibitem [\protect \citeauthoryear {%
Collins%
\ \protect \BOthers {.}}{%
Collins%
\ \protect \BOthers {.}}{%
{\protect \APACyear {2022}}%
}]{%
collins2022predicting}
\APACinsertmetastar {%
collins2022predicting}%
\begin{APACrefauthors}%
Collins, E\BPBI L.%
, Sanchez, G\BPBI M.%
, Terando, A.%
, Stillwell, C\BPBI C.%
, Mitasova, H.%
, Sebastian, A.%
\BCBL {}\ \BBA {} Meentemeyer, R\BPBI K.%
\end{APACrefauthors}%
\unskip\
\newblock
\APACrefYearMonthDay{2022}{}{}.
\newblock
{\BBOQ}\APACrefatitle {Predicting flood damage probability across the
  conterminous United States} {Predicting flood damage probability across the
  conterminous united states}.{\BBCQ}
\newblock
\APACjournalVolNumPages{Environmental Research Letters}{17}{3}{034006}.
\PrintBackRefs{\CurrentBib}

\bibitem [\protect \citeauthoryear {%
Cornwall%
}{%
Cornwall%
}{%
{\protect \APACyear {2021}}%
}]{%
cornwall2021europe}
\APACinsertmetastar {%
cornwall2021europe}%
\begin{APACrefauthors}%
Cornwall, W.%
\end{APACrefauthors}%
\unskip\
\newblock
\APACrefYearMonthDay{2021}{}{}.
\newblock
\APACrefbtitle {Europe's deadly floods leave scientists stunned.} {Europe's
  deadly floods leave scientists stunned.}
\newblock
\APACaddressPublisher{}{American Association for the Advancement of Science}.
\PrintBackRefs{\CurrentBib}

\bibitem [\protect \citeauthoryear {%
Cox%
}{%
Cox%
}{%
{\protect \APACyear {1958}}%
}]{%
cox1958regression}
\APACinsertmetastar {%
cox1958regression}%
\begin{APACrefauthors}%
Cox, D\BPBI R.%
\end{APACrefauthors}%
\unskip\
\newblock
\APACrefYearMonthDay{1958}{}{}.
\newblock
{\BBOQ}\APACrefatitle {The regression analysis of binary sequences} {The
  regression analysis of binary sequences}.{\BBCQ}
\newblock
\APACjournalVolNumPages{Journal of the Royal Statistical Society Series B:
  Statistical Methodology}{20}{2}{215--232}.
\PrintBackRefs{\CurrentBib}

\bibitem [\protect \citeauthoryear {%
D'Odorico%
\ \BBA {} Rigon%
}{%
D'Odorico%
\ \BBA {} Rigon%
}{%
{\protect \APACyear {2003}}%
}]{%
odorico2003hillslope}
\APACinsertmetastar {%
odorico2003hillslope}%
\begin{APACrefauthors}%
D'Odorico, P.%
\BCBT {}\ \BBA {} Rigon, R.%
\end{APACrefauthors}%
\unskip\
\newblock
\APACrefYearMonthDay{2003}{}{}.
\newblock
{\BBOQ}\APACrefatitle {Hillslope and channel contributions to the hydrologic
  response} {Hillslope and channel contributions to the hydrologic
  response}.{\BBCQ}
\newblock
\APACjournalVolNumPages{Water resources research}{39}{5}{}.
\PrintBackRefs{\CurrentBib}

\bibitem [\protect \citeauthoryear {%
Dowell%
\ \protect \BOthers {.}}{%
Dowell%
\ \protect \BOthers {.}}{%
{\protect \APACyear {2022}}%
}]{%
dowell2022high}
\APACinsertmetastar {%
dowell2022high}%
\begin{APACrefauthors}%
Dowell, D\BPBI C.%
, Alexander, C\BPBI R.%
, James, E\BPBI P.%
, Weygandt, S\BPBI S.%
, Benjamin, S\BPBI G.%
, Manikin, G\BPBI S.%
\BDBL {}others%
\end{APACrefauthors}%
\unskip\
\newblock
\APACrefYearMonthDay{2022}{}{}.
\newblock
{\BBOQ}\APACrefatitle {The High-Resolution Rapid Refresh (HRRR): An hourly
  updating convection-allowing forecast model. Part I: Motivation and system
  description} {The high-resolution rapid refresh (hrrr): An hourly updating
  convection-allowing forecast model. part i: Motivation and system
  description}.{\BBCQ}
\newblock
\APACjournalVolNumPages{Weather and Forecasting}{37}{8}{1371--1395}.
\PrintBackRefs{\CurrentBib}

\bibitem [\protect \citeauthoryear {%
Ducharne%
}{%
Ducharne%
}{%
{\protect \APACyear {2009}}%
}]{%
ducharne2009reducing}
\APACinsertmetastar {%
ducharne2009reducing}%
\begin{APACrefauthors}%
Ducharne, A.%
\end{APACrefauthors}%
\unskip\
\newblock
\APACrefYearMonthDay{2009}{}{}.
\newblock
{\BBOQ}\APACrefatitle {Reducing scale dependence in TOPMODEL using a
  dimensionless topographic index} {Reducing scale dependence in topmodel using
  a dimensionless topographic index}.{\BBCQ}
\newblock
\APACjournalVolNumPages{Hydrology and Earth System
  Sciences}{13}{12}{2399--2412}.
\PrintBackRefs{\CurrentBib}

\bibitem [\protect \citeauthoryear {%
Ercan%
, Kavvas%
\BCBL {}\ \BBA {} Haltas%
}{%
Ercan%
\ \protect \BOthers {.}}{%
{\protect \APACyear {2014}}%
}]{%
ercan2014scaling}
\APACinsertmetastar {%
ercan2014scaling}%
\begin{APACrefauthors}%
Ercan, A.%
, Kavvas, M\BPBI L.%
\BCBL {}\ \BBA {} Haltas, I.%
\end{APACrefauthors}%
\unskip\
\newblock
\APACrefYearMonthDay{2014}{}{}.
\newblock
{\BBOQ}\APACrefatitle {Scaling and self-similarity in one-dimensional unsteady
  open channel flow} {Scaling and self-similarity in one-dimensional unsteady
  open channel flow}.{\BBCQ}
\newblock
\APACjournalVolNumPages{Hydrological Processes}{28}{5}{2721--2737}.
\PrintBackRefs{\CurrentBib}

\bibitem [\protect \citeauthoryear {%
{Federal Emergency Management Agency}%
}{%
{Federal Emergency Management Agency}%
}{%
{\protect \APACyear {2024}}%
}]{%
fema}
\APACinsertmetastar {%
fema}%
\begin{APACrefauthors}%
{Federal Emergency Management Agency}.%
\end{APACrefauthors}%
\unskip\
\newblock
\APACrefYearMonthDay{2024}{}{}.
\newblock
\APACrefbtitle {{FEMA} {M}ap {S}ervice {C}enter [{D}ataset].} {{FEMA} {M}ap
  {S}ervice {C}enter [{D}ataset].}
\newblock
\begin{APACrefURL} \url{https://msc.fema.gov/portal/home} \end{APACrefURL}
\newblock
\APACrefnote{Accessed: 2024-11-01}
\PrintBackRefs{\CurrentBib}

\bibitem [\protect \citeauthoryear {%
Garousi-Nejad%
, Tarboton%
, Aboutalebi%
\BCBL {}\ \BBA {} Torres-Rua%
}{%
Garousi-Nejad%
\ \protect \BOthers {.}}{%
{\protect \APACyear {2019}}%
}]{%
garousi2019terrain}
\APACinsertmetastar {%
garousi2019terrain}%
\begin{APACrefauthors}%
Garousi-Nejad, I.%
, Tarboton, D\BPBI G.%
, Aboutalebi, M.%
\BCBL {}\ \BBA {} Torres-Rua, A\BPBI F.%
\end{APACrefauthors}%
\unskip\
\newblock
\APACrefYearMonthDay{2019}{}{}.
\newblock
{\BBOQ}\APACrefatitle {Terrain analysis enhancements to the height above
  nearest drainage flood inundation mapping method} {Terrain analysis
  enhancements to the height above nearest drainage flood inundation mapping
  method}.{\BBCQ}
\newblock
\APACjournalVolNumPages{Water Resources Research}{55}{10}{7983--8009}.
\PrintBackRefs{\CurrentBib}

\bibitem [\protect \citeauthoryear {%
Giovannettone%
, Copenhaver%
, Burns%
\BCBL {}\ \BBA {} Choquette%
}{%
Giovannettone%
\ \protect \BOthers {.}}{%
{\protect \APACyear {2018}}%
}]{%
giovannettone2018statistical}
\APACinsertmetastar {%
giovannettone2018statistical}%
\begin{APACrefauthors}%
Giovannettone, J.%
, Copenhaver, T.%
, Burns, M.%
\BCBL {}\ \BBA {} Choquette, S.%
\end{APACrefauthors}%
\unskip\
\newblock
\APACrefYearMonthDay{2018}{}{}.
\newblock
{\BBOQ}\APACrefatitle {A statistical approach to mapping flood susceptibility
  in the Lower Connecticut River Valley Region} {A statistical approach to
  mapping flood susceptibility in the lower connecticut river valley
  region}.{\BBCQ}
\newblock
\APACjournalVolNumPages{Water Resources Research}{54}{10}{7603--7618}.
\PrintBackRefs{\CurrentBib}

\bibitem [\protect \citeauthoryear {%
{Government Accountability Office}%
}{%
{Government Accountability Office}%
}{%
{\protect \APACyear {2021}}%
}]{%
gao2021flooding}
\APACinsertmetastar {%
gao2021flooding}%
\begin{APACrefauthors}%
{Government Accountability Office}.%
\end{APACrefauthors}%
\unskip\
\newblock
\APACrefYearMonthDay{2021}{}{}.
\newblock
\APACrefbtitle {Better Planning and Analysis Needed to Address Current and
  Future Flood Hazards} {Better planning and analysis needed to address current
  and future flood hazards}\ \APACbVolEdTR{}{\BTR{}\ \BNUM\ GAO-22-104079}.
\newblock
\APACaddressInstitution{Washington, DC}{United States Government Accountability
  Office}.
\PrintBackRefs{\CurrentBib}

\bibitem [\protect \citeauthoryear {%
{GRASS Development Team}%
}{%
{GRASS Development Team}%
}{%
{\protect \APACyear {2017}}%
}]{%
GRASS_GIS_software}
\APACinsertmetastar {%
GRASS_GIS_software}%
\begin{APACrefauthors}%
{GRASS Development Team}.%
\end{APACrefauthors}%
\unskip\
\newblock
\APACrefYearMonthDay{2017}{}{}.
\newblock
\APACrefbtitle {Geographic {R}esources {A}nalysis {S}upport {S}ystem ({GRASS
  GIS}) {S}oftware, {V}ersion 7.2 [{S}oftware].} {Geographic {R}esources
  {A}nalysis {S}upport {S}ystem ({GRASS GIS}) {S}oftware, {V}ersion 7.2
  [{S}oftware].}
\newblock
\begin{APACrefURL} \url{http://grass.osgeo.org} \end{APACrefURL}
\PrintBackRefs{\CurrentBib}

\bibitem [\protect \citeauthoryear {%
Gunaratnam%
, Degroff%
\BCBL {}\ \BBA {} Gero%
}{%
Gunaratnam%
\ \protect \BOthers {.}}{%
{\protect \APACyear {2003}}%
}]{%
gunaratnam2003improving}
\APACinsertmetastar {%
gunaratnam2003improving}%
\begin{APACrefauthors}%
Gunaratnam, D\BPBI J.%
, Degroff, T.%
\BCBL {}\ \BBA {} Gero, J\BPBI S.%
\end{APACrefauthors}%
\unskip\
\newblock
\APACrefYearMonthDay{2003}{}{}.
\newblock
{\BBOQ}\APACrefatitle {Improving neural network models of physical systems
  through dimensional analysis} {Improving neural network models of physical
  systems through dimensional analysis}.{\BBCQ}
\newblock
\APACjournalVolNumPages{Applied Soft Computing}{2}{4}{283--296}.
\PrintBackRefs{\CurrentBib}

\bibitem [\protect \citeauthoryear {%
Guo%
, Moosavi%
\BCBL {}\ \BBA {} Leit{\~a}o%
}{%
Guo%
\ \protect \BOthers {.}}{%
{\protect \APACyear {2022}}%
}]{%
guo2022data}
\APACinsertmetastar {%
guo2022data}%
\begin{APACrefauthors}%
Guo, Z.%
, Moosavi, V.%
\BCBL {}\ \BBA {} Leit{\~a}o, J\BPBI P.%
\end{APACrefauthors}%
\unskip\
\newblock
\APACrefYearMonthDay{2022}{}{}.
\newblock
{\BBOQ}\APACrefatitle {Data-driven rapid flood prediction mapping with
  catchment generalizability} {Data-driven rapid flood prediction mapping with
  catchment generalizability}.{\BBCQ}
\newblock
\APACjournalVolNumPages{Journal of Hydrology}{609}{}{127726}.
\PrintBackRefs{\CurrentBib}

\bibitem [\protect \citeauthoryear {%
Hocini%
\ \protect \BOthers {.}}{%
Hocini%
\ \protect \BOthers {.}}{%
{\protect \APACyear {2021}}%
}]{%
hocini2021performance}
\APACinsertmetastar {%
hocini2021performance}%
\begin{APACrefauthors}%
Hocini, N.%
, Payrastre, O.%
, Bourgin, F.%
, Gaume, E.%
, Davy, P.%
, Lague, D.%
\BDBL {}Pons, F.%
\end{APACrefauthors}%
\unskip\
\newblock
\APACrefYearMonthDay{2021}{}{}.
\newblock
{\BBOQ}\APACrefatitle {Performance of automated methods for flash flood
  inundation mapping: a comparison of a digital terrain model (DTM) filling and
  two hydrodynamic methods} {Performance of automated methods for flash flood
  inundation mapping: a comparison of a digital terrain model (dtm) filling and
  two hydrodynamic methods}.{\BBCQ}
\newblock
\APACjournalVolNumPages{Hydrology and Earth System
  Sciences}{25}{6}{2979--2995}.
\PrintBackRefs{\CurrentBib}

\bibitem [\protect \citeauthoryear {%
Hosmer~Jr%
, Lemeshow%
\BCBL {}\ \BBA {} Sturdivant%
}{%
Hosmer~Jr%
\ \protect \BOthers {.}}{%
{\protect \APACyear {2013}}%
}]{%
hosmer2013applied}
\APACinsertmetastar {%
hosmer2013applied}%
\begin{APACrefauthors}%
Hosmer~Jr, D\BPBI W.%
, Lemeshow, S.%
\BCBL {}\ \BBA {} Sturdivant, R\BPBI X.%
\end{APACrefauthors}%
\unskip\
\newblock
\APACrefYear{2013}.
\newblock
\APACrefbtitle {Applied logistic regression} {Applied logistic regression}\
  (\BVOL~398).
\newblock
\APACaddressPublisher{}{John Wiley \& Sons}.
\PrintBackRefs{\CurrentBib}

\bibitem [\protect \citeauthoryear {%
Hosseiny%
, Nazari%
, Smith%
\BCBL {}\ \BBA {} Nataraj%
}{%
Hosseiny%
\ \protect \BOthers {.}}{%
{\protect \APACyear {2020}}%
}]{%
hosseiny2020framework}
\APACinsertmetastar {%
hosseiny2020framework}%
\begin{APACrefauthors}%
Hosseiny, H.%
, Nazari, F.%
, Smith, V.%
\BCBL {}\ \BBA {} Nataraj, C.%
\end{APACrefauthors}%
\unskip\
\newblock
\APACrefYearMonthDay{2020}{}{}.
\newblock
{\BBOQ}\APACrefatitle {A framework for modeling flood depth using a hybrid of
  hydraulics and machine learning} {A framework for modeling flood depth using
  a hybrid of hydraulics and machine learning}.{\BBCQ}
\newblock
\APACjournalVolNumPages{Scientific Reports}{10}{1}{1--14}.
\PrintBackRefs{\CurrentBib}

\bibitem [\protect \citeauthoryear {%
Hu%
, Fang%
, Pain%
\BCBL {}\ \BBA {} Navon%
}{%
Hu%
\ \protect \BOthers {.}}{%
{\protect \APACyear {2019}}%
}]{%
hu2019rapid}
\APACinsertmetastar {%
hu2019rapid}%
\begin{APACrefauthors}%
Hu, R.%
, Fang, F.%
, Pain, C.%
\BCBL {}\ \BBA {} Navon, I.%
\end{APACrefauthors}%
\unskip\
\newblock
\APACrefYearMonthDay{2019}{}{}.
\newblock
{\BBOQ}\APACrefatitle {Rapid spatio-temporal flood prediction and uncertainty
  quantification using a deep learning method} {Rapid spatio-temporal flood
  prediction and uncertainty quantification using a deep learning
  method}.{\BBCQ}
\newblock
\APACjournalVolNumPages{Journal of Hydrology}{575}{}{911--920}.
\PrintBackRefs{\CurrentBib}

\bibitem [\protect \citeauthoryear {%
Ivanov%
\ \protect \BOthers {.}}{%
Ivanov%
\ \protect \BOthers {.}}{%
{\protect \APACyear {2021}}%
}]{%
ivanov2021breaking}
\APACinsertmetastar {%
ivanov2021breaking}%
\begin{APACrefauthors}%
Ivanov, V\BPBI Y.%
, Xu, D.%
, Dwelle, M\BPBI C.%
, Sargsyan, K.%
, Wright, D\BPBI B.%
, Katopodes, N.%
\BDBL {}others%
\end{APACrefauthors}%
\unskip\
\newblock
\APACrefYearMonthDay{2021}{}{}.
\newblock
{\BBOQ}\APACrefatitle {Breaking down the computational barriers to real-time
  urban flood forecasting} {Breaking down the computational barriers to
  real-time urban flood forecasting}.{\BBCQ}
\newblock
\APACjournalVolNumPages{Geophysical Research Letters}{48}{20}{e2021GL093585}.
\PrintBackRefs{\CurrentBib}

\bibitem [\protect \citeauthoryear {%
Jakeman%
\ \BBA {} Hornberger%
}{%
Jakeman%
\ \BBA {} Hornberger%
}{%
{\protect \APACyear {1993}}%
}]{%
jakeman1993much}
\APACinsertmetastar {%
jakeman1993much}%
\begin{APACrefauthors}%
Jakeman, A.%
\BCBT {}\ \BBA {} Hornberger, G.%
\end{APACrefauthors}%
\unskip\
\newblock
\APACrefYearMonthDay{1993}{}{}.
\newblock
{\BBOQ}\APACrefatitle {How much complexity is warranted in a rainfall-runoff
  model?} {How much complexity is warranted in a rainfall-runoff model?}{\BBCQ}
\newblock
\APACjournalVolNumPages{Water Resources Research}{29}{8}{2637--2649}.
\PrintBackRefs{\CurrentBib}

\bibitem [\protect \citeauthoryear {%
Jasiewicz%
\ \BBA {} Metz%
}{%
Jasiewicz%
\ \BBA {} Metz%
}{%
{\protect \APACyear {2011}}%
}]{%
jasiewicz2011new}
\APACinsertmetastar {%
jasiewicz2011new}%
\begin{APACrefauthors}%
Jasiewicz, J.%
\BCBT {}\ \BBA {} Metz, M.%
\end{APACrefauthors}%
\unskip\
\newblock
\APACrefYearMonthDay{2011}{}{}.
\newblock
{\BBOQ}\APACrefatitle {A new {GRASS GIS} toolkit for Hortonian analysis of
  drainage networks} {A new {GRASS GIS} toolkit for hortonian analysis of
  drainage networks}.{\BBCQ}
\newblock
\APACjournalVolNumPages{Computers \& Geosciences}{37}{8}{1162--1173}.
\PrintBackRefs{\CurrentBib}

\bibitem [\protect \citeauthoryear {%
Johnson%
, Munasinghe%
, Eyelade%
\BCBL {}\ \BBA {} Cohen%
}{%
Johnson%
\ \protect \BOthers {.}}{%
{\protect \APACyear {2019}}%
}]{%
johnson2019integrated}
\APACinsertmetastar {%
johnson2019integrated}%
\begin{APACrefauthors}%
Johnson, J\BPBI M.%
, Munasinghe, D.%
, Eyelade, D.%
\BCBL {}\ \BBA {} Cohen, S.%
\end{APACrefauthors}%
\unskip\
\newblock
\APACrefYearMonthDay{2019}{}{}.
\newblock
{\BBOQ}\APACrefatitle {An integrated evaluation of the national water model
  (NWM)--Height above nearest drainage (HAND) flood mapping methodology} {An
  integrated evaluation of the national water model (nwm)--height above nearest
  drainage (hand) flood mapping methodology}.{\BBCQ}
\newblock
\APACjournalVolNumPages{Natural Hazards and Earth System
  Sciences}{19}{11}{2405--2420}.
\PrintBackRefs{\CurrentBib}

\bibitem [\protect \citeauthoryear {%
Maidment%
, Olivera%
, Calver%
, Eatherall%
\BCBL {}\ \BBA {} Fraczek%
}{%
Maidment%
\ \protect \BOthers {.}}{%
{\protect \APACyear {1996}}%
}]{%
maidment1996unit}
\APACinsertmetastar {%
maidment1996unit}%
\begin{APACrefauthors}%
Maidment, D.%
, Olivera, F.%
, Calver, A.%
, Eatherall, A.%
\BCBL {}\ \BBA {} Fraczek, W.%
\end{APACrefauthors}%
\unskip\
\newblock
\APACrefYearMonthDay{1996}{}{}.
\newblock
{\BBOQ}\APACrefatitle {Unit hydrograph derived from a spatially distributed
  velocity field} {Unit hydrograph derived from a spatially distributed
  velocity field}.{\BBCQ}
\newblock
\APACjournalVolNumPages{Hydrological processes}{10}{6}{831--844}.
\PrintBackRefs{\CurrentBib}

\bibitem [\protect \citeauthoryear {%
Manfreda%
, Nardi%
\BCBL {}\ \protect \BOthers {.}}{%
Manfreda%
, Nardi%
\BCBL {}\ \protect \BOthers {.}}{%
{\protect \APACyear {2014}}%
}]{%
manfreda2014investigation}
\APACinsertmetastar {%
manfreda2014investigation}%
\begin{APACrefauthors}%
Manfreda, S.%
, Nardi, F.%
, Samela, C.%
, Grimaldi, S.%
, Taramasso, A\BPBI C.%
, Roth, G.%
\BCBL {}\ \BBA {} Sole, A.%
\end{APACrefauthors}%
\unskip\
\newblock
\APACrefYearMonthDay{2014}{}{}.
\newblock
{\BBOQ}\APACrefatitle {Investigation on the use of geomorphic approaches for
  the delineation of flood prone areas} {Investigation on the use of geomorphic
  approaches for the delineation of flood prone areas}.{\BBCQ}
\newblock
\APACjournalVolNumPages{Journal of hydrology}{517}{}{863--876}.
\PrintBackRefs{\CurrentBib}

\bibitem [\protect \citeauthoryear {%
Manfreda%
, Samela%
, Sole%
\BCBL {}\ \BBA {} Fiorentino%
}{%
Manfreda%
, Samela%
\BCBL {}\ \protect \BOthers {.}}{%
{\protect \APACyear {2014}}%
}]{%
manfreda2014flood}
\APACinsertmetastar {%
manfreda2014flood}%
\begin{APACrefauthors}%
Manfreda, S.%
, Samela, C.%
, Sole, A.%
\BCBL {}\ \BBA {} Fiorentino, M.%
\end{APACrefauthors}%
\unskip\
\newblock
\APACrefYearMonthDay{2014}{}{}.
\newblock
{\BBOQ}\APACrefatitle {Flood-prone areas assessment using linear binary
  classifiers based on morphological indices} {Flood-prone areas assessment
  using linear binary classifiers based on morphological indices}.{\BBCQ}
\newblock
\BIn{} \APACrefbtitle {Vulnerability, Uncertainty, and Risk: Quantification,
  Mitigation, and Management} {Vulnerability, uncertainty, and risk:
  Quantification, mitigation, and management}\ (\BPGS\ 2002--2011).
\PrintBackRefs{\CurrentBib}

\bibitem [\protect \citeauthoryear {%
Meles%
, Younger%
, Jackson%
, Du%
\BCBL {}\ \BBA {} Drover%
}{%
Meles%
\ \protect \BOthers {.}}{%
{\protect \APACyear {2020}}%
}]{%
meles2020wetness}
\APACinsertmetastar {%
meles2020wetness}%
\begin{APACrefauthors}%
Meles, M\BPBI B.%
, Younger, S\BPBI E.%
, Jackson, C\BPBI R.%
, Du, E.%
\BCBL {}\ \BBA {} Drover, D.%
\end{APACrefauthors}%
\unskip\
\newblock
\APACrefYearMonthDay{2020}{}{}.
\newblock
{\BBOQ}\APACrefatitle {Wetness index based on landscape position and topography
  (WILT): Modifying TWI to reflect landscape position} {Wetness index based on
  landscape position and topography (wilt): Modifying twi to reflect landscape
  position}.{\BBCQ}
\newblock
\APACjournalVolNumPages{Journal of environmental management}{255}{}{109863}.
\PrintBackRefs{\CurrentBib}

\bibitem [\protect \citeauthoryear {%
Moussa%
\ \BBA {} Bocquillon%
}{%
Moussa%
\ \BBA {} Bocquillon%
}{%
{\protect \APACyear {2000}}%
}]{%
moussa2000approximation}
\APACinsertmetastar {%
moussa2000approximation}%
\begin{APACrefauthors}%
Moussa, R.%
\BCBT {}\ \BBA {} Bocquillon, C.%
\end{APACrefauthors}%
\unskip\
\newblock
\APACrefYearMonthDay{2000}{}{}.
\newblock
{\BBOQ}\APACrefatitle {Approximation zones of the Saint-Venant equations f
  flood routing with overbank flow} {Approximation zones of the saint-venant
  equations f flood routing with overbank flow}.{\BBCQ}
\newblock
\APACjournalVolNumPages{Hydrology and Earth System Sciences}{4}{2}{251--260}.
\PrintBackRefs{\CurrentBib}

\bibitem [\protect \citeauthoryear {%
Murphy%
}{%
Murphy%
}{%
{\protect \APACyear {2022}}%
}]{%
murphy2022probabilistic}
\APACinsertmetastar {%
murphy2022probabilistic}%
\begin{APACrefauthors}%
Murphy, K\BPBI P.%
\end{APACrefauthors}%
\unskip\
\newblock
\APACrefYear{2022}.
\newblock
\APACrefbtitle {Probabilistic machine learning: an introduction} {Probabilistic
  machine learning: an introduction}.
\newblock
\APACaddressPublisher{}{MIT press}.
\PrintBackRefs{\CurrentBib}

\bibitem [\protect \citeauthoryear {%
Nachappa%
\ \protect \BOthers {.}}{%
Nachappa%
\ \protect \BOthers {.}}{%
{\protect \APACyear {2020}}%
}]{%
nachappa2020flood}
\APACinsertmetastar {%
nachappa2020flood}%
\begin{APACrefauthors}%
Nachappa, T\BPBI G.%
, Piralilou, S\BPBI T.%
, Gholamnia, K.%
, Ghorbanzadeh, O.%
, Rahmati, O.%
\BCBL {}\ \BBA {} Blaschke, T.%
\end{APACrefauthors}%
\unskip\
\newblock
\APACrefYearMonthDay{2020}{}{}.
\newblock
{\BBOQ}\APACrefatitle {Flood susceptibility mapping with machine learning,
  multi-criteria decision analysis and ensemble using Dempster Shafer Theory}
  {Flood susceptibility mapping with machine learning, multi-criteria decision
  analysis and ensemble using dempster shafer theory}.{\BBCQ}
\newblock
\APACjournalVolNumPages{Journal of Hydrology}{590}{}{125275}.
\PrintBackRefs{\CurrentBib}

\bibitem [\protect \citeauthoryear {%
Nardi%
, Vivoni%
\BCBL {}\ \BBA {} Grimaldi%
}{%
Nardi%
\ \protect \BOthers {.}}{%
{\protect \APACyear {2006}}%
}]{%
nardi2006investigating}
\APACinsertmetastar {%
nardi2006investigating}%
\begin{APACrefauthors}%
Nardi, F.%
, Vivoni, E\BPBI R.%
\BCBL {}\ \BBA {} Grimaldi, S.%
\end{APACrefauthors}%
\unskip\
\newblock
\APACrefYearMonthDay{2006}{}{}.
\newblock
{\BBOQ}\APACrefatitle {Investigating a floodplain scaling relation using a
  hydrogeomorphic delineation method} {Investigating a floodplain scaling
  relation using a hydrogeomorphic delineation method}.{\BBCQ}
\newblock
\APACjournalVolNumPages{Water Resources Research}{42}{9}{}.
\PrintBackRefs{\CurrentBib}

\bibitem [\protect \citeauthoryear {%
Oppenheimer%
, Doman%
\BCBL {}\ \BBA {} Merrick%
}{%
Oppenheimer%
\ \protect \BOthers {.}}{%
{\protect \APACyear {2023}}%
}]{%
oppenheimer2023multi}
\APACinsertmetastar {%
oppenheimer2023multi}%
\begin{APACrefauthors}%
Oppenheimer, M\BPBI W.%
, Doman, D\BPBI B.%
\BCBL {}\ \BBA {} Merrick, J\BPBI D.%
\end{APACrefauthors}%
\unskip\
\newblock
\APACrefYearMonthDay{2023}{}{}.
\newblock
{\BBOQ}\APACrefatitle {Multi-scale physics-informed machine learning using the
  Buckingham Pi theorem} {Multi-scale physics-informed machine learning using
  the buckingham pi theorem}.{\BBCQ}
\newblock
\APACjournalVolNumPages{Journal of Computational Physics}{474}{}{111810}.
\PrintBackRefs{\CurrentBib}

\bibitem [\protect \citeauthoryear {%
Pakdehi%
, Ahmadisharaf%
, Nazari%
\BCBL {}\ \BBA {} Cho%
}{%
Pakdehi%
\ \protect \BOthers {.}}{%
{\protect \APACyear {2023}}%
}]{%
pakdehi2023transferability}
\APACinsertmetastar {%
pakdehi2023transferability}%
\begin{APACrefauthors}%
Pakdehi, M.%
, Ahmadisharaf, E.%
, Nazari, B.%
\BCBL {}\ \BBA {} Cho, E.%
\end{APACrefauthors}%
\unskip\
\newblock
\APACrefYearMonthDay{2023}{}{}.
\newblock
{\BBOQ}\APACrefatitle {Transferability of machine learning-based modeling
  frameworks across flood events for hindcasting maximum river flood depths in
  coastal watersheds} {Transferability of machine learning-based modeling
  frameworks across flood events for hindcasting maximum river flood depths in
  coastal watersheds}.{\BBCQ}
\newblock
\APACjournalVolNumPages{Natural Hazards and Earth System Sciences
  Discussions}{2023}{}{1--57}.
\PrintBackRefs{\CurrentBib}

\bibitem [\protect \citeauthoryear {%
Pedregosa%
\ \protect \BOthers {.}}{%
Pedregosa%
\ \protect \BOthers {.}}{%
{\protect \APACyear {2011}}%
}]{%
scikit-learn}
\APACinsertmetastar {%
scikit-learn}%
\begin{APACrefauthors}%
Pedregosa, F.%
, Varoquaux, G.%
, Gramfort, A.%
, Michel, V.%
, Thirion, B.%
, Grisel, O.%
\BDBL {}Duchesnay, E.%
\end{APACrefauthors}%
\unskip\
\newblock
\APACrefYearMonthDay{2011}{}{}.
\newblock
{\BBOQ}\APACrefatitle {Scikit-learn: Machine Learning in {P}ython [{S}oftware]}
  {Scikit-learn: Machine learning in {P}ython [{S}oftware]}.{\BBCQ}
\newblock
\APACjournalVolNumPages{Journal of Machine Learning
  Research}{12}{}{2825--2830}.
\newblock
\begin{APACrefURL} \url{https://scikit-learn.org/stable/} \end{APACrefURL}
\PrintBackRefs{\CurrentBib}

\bibitem [\protect \citeauthoryear {%
Porporato%
}{%
Porporato%
}{%
{\protect \APACyear {2022}}%
}]{%
porporato2022hydrology}
\APACinsertmetastar {%
porporato2022hydrology}%
\begin{APACrefauthors}%
Porporato, A.%
\end{APACrefauthors}%
\unskip\
\newblock
\APACrefYearMonthDay{2022}{}{}.
\newblock
{\BBOQ}\APACrefatitle {Hydrology without dimensions} {Hydrology without
  dimensions}.{\BBCQ}
\newblock
\APACjournalVolNumPages{Hydrology and Earth System Sciences}{26}{2}{355--374}.
\PrintBackRefs{\CurrentBib}

\bibitem [\protect \citeauthoryear {%
Porporato%
\ \BBA {} Yin%
}{%
Porporato%
\ \BBA {} Yin%
}{%
{\protect \APACyear {2022}}%
}]{%
porporato2022ecohydrology}
\APACinsertmetastar {%
porporato2022ecohydrology}%
\begin{APACrefauthors}%
Porporato, A.%
\BCBT {}\ \BBA {} Yin, J.%
\end{APACrefauthors}%
\unskip\
\newblock
\APACrefYear{2022}.
\newblock
\APACrefbtitle {Ecohydrology: Dynamics of Life and Water in the Critical Zone}
  {Ecohydrology: Dynamics of life and water in the critical zone}.
\newblock
\APACaddressPublisher{}{Cambridge University Press}.
\newblock
\begin{APACrefURL} \url{https://books.google.com/books?id=-3J-zgEACAAJ}
  \end{APACrefURL}
\PrintBackRefs{\CurrentBib}

\bibitem [\protect \citeauthoryear {%
Rahmati%
\ \protect \BOthers {.}}{%
Rahmati%
\ \protect \BOthers {.}}{%
{\protect \APACyear {2020}}%
}]{%
rahmati2020development}
\APACinsertmetastar {%
rahmati2020development}%
\begin{APACrefauthors}%
Rahmati, O.%
, Darabi, H.%
, Panahi, M.%
, Kalantari, Z.%
, Naghibi, S\BPBI A.%
, Ferreira, C\BPBI S\BPBI S.%
\BDBL {}others%
\end{APACrefauthors}%
\unskip\
\newblock
\APACrefYearMonthDay{2020}{}{}.
\newblock
{\BBOQ}\APACrefatitle {Development of novel hybridized models for urban flood
  susceptibility mapping} {Development of novel hybridized models for urban
  flood susceptibility mapping}.{\BBCQ}
\newblock
\APACjournalVolNumPages{Scientific reports}{10}{1}{1--19}.
\PrintBackRefs{\CurrentBib}

\bibitem [\protect \citeauthoryear {%
Rigon%
, D'Odorico%
\BCBL {}\ \BBA {} Bertoldi%
}{%
Rigon%
\ \protect \BOthers {.}}{%
{\protect \APACyear {2011}}%
}]{%
rigon2011geomorphic}
\APACinsertmetastar {%
rigon2011geomorphic}%
\begin{APACrefauthors}%
Rigon, R.%
, D'Odorico, P.%
\BCBL {}\ \BBA {} Bertoldi, G.%
\end{APACrefauthors}%
\unskip\
\newblock
\APACrefYearMonthDay{2011}{}{}.
\newblock
{\BBOQ}\APACrefatitle {The geomorphic structure of the runoff peak} {The
  geomorphic structure of the runoff peak}.{\BBCQ}
\newblock
\APACjournalVolNumPages{Hydrology and Earth System
  Sciences}{15}{6}{1853--1863}.
\PrintBackRefs{\CurrentBib}

\bibitem [\protect \citeauthoryear {%
Rinaldo%
, Marani%
\BCBL {}\ \BBA {} Rigon%
}{%
Rinaldo%
\ \protect \BOthers {.}}{%
{\protect \APACyear {1991}}%
}]{%
rinaldo1991geomorphological}
\APACinsertmetastar {%
rinaldo1991geomorphological}%
\begin{APACrefauthors}%
Rinaldo, A.%
, Marani, A.%
\BCBL {}\ \BBA {} Rigon, R.%
\end{APACrefauthors}%
\unskip\
\newblock
\APACrefYearMonthDay{1991}{}{}.
\newblock
{\BBOQ}\APACrefatitle {Geomorphological dispersion} {Geomorphological
  dispersion}.{\BBCQ}
\newblock
\APACjournalVolNumPages{Water Resources Research}{27}{4}{513--525}.
\PrintBackRefs{\CurrentBib}

\bibitem [\protect \citeauthoryear {%
Rodr{\'\i}guez-Iturbe%
\ \BBA {} Porporato%
}{%
Rodr{\'\i}guez-Iturbe%
\ \BBA {} Porporato%
}{%
{\protect \APACyear {2004}}%
}]{%
rodrigueziturbe2004ecohydrology}
\APACinsertmetastar {%
rodrigueziturbe2004ecohydrology}%
\begin{APACrefauthors}%
Rodr{\'\i}guez-Iturbe, I.%
\BCBT {}\ \BBA {} Porporato, A.%
\end{APACrefauthors}%
\unskip\
\newblock
\APACrefYear{2004}.
\newblock
\APACrefbtitle {Ecohydrology of water-controlled ecosystems: soil moisture and
  plant dynamics} {Ecohydrology of water-controlled ecosystems: soil moisture
  and plant dynamics}.
\newblock
\APACaddressPublisher{}{Cambridge University Press}.
\PrintBackRefs{\CurrentBib}

\bibitem [\protect \citeauthoryear {%
Ronneberger%
, Fischer%
\BCBL {}\ \BBA {} Brox%
}{%
Ronneberger%
\ \protect \BOthers {.}}{%
{\protect \APACyear {2015}}%
}]{%
ronneberger2015u}
\APACinsertmetastar {%
ronneberger2015u}%
\begin{APACrefauthors}%
Ronneberger, O.%
, Fischer, P.%
\BCBL {}\ \BBA {} Brox, T.%
\end{APACrefauthors}%
\unskip\
\newblock
\APACrefYearMonthDay{2015}{}{}.
\newblock
{\BBOQ}\APACrefatitle {U-net: Convolutional networks for biomedical image
  segmentation} {U-net: Convolutional networks for biomedical image
  segmentation}.{\BBCQ}
\newblock
\BIn{} \APACrefbtitle {Medical image computing and computer-assisted
  intervention--MICCAI 2015: 18th international conference, Munich, Germany,
  October 5-9, 2015, proceedings, part III 18} {Medical image computing and
  computer-assisted intervention--miccai 2015: 18th international conference,
  munich, germany, october 5-9, 2015, proceedings, part iii 18}\ (\BPGS\
  234--241).
\PrintBackRefs{\CurrentBib}

\bibitem [\protect \citeauthoryear {%
Rudolph%
}{%
Rudolph%
}{%
{\protect \APACyear {1998}}%
}]{%
rudolph1998context}
\APACinsertmetastar {%
rudolph1998context}%
\begin{APACrefauthors}%
Rudolph, S.%
\end{APACrefauthors}%
\unskip\
\newblock
\APACrefYearMonthDay{1998}{}{}.
\newblock
{\BBOQ}\APACrefatitle {On the context of dimensional analysis in artificial
  intelligence} {On the context of dimensional analysis in artificial
  intelligence}.{\BBCQ}
\newblock
\BIn{} \APACrefbtitle {International Workshop on Similarity Methods.}
  {International workshop on similarity methods.}
\PrintBackRefs{\CurrentBib}

\bibitem [\protect \citeauthoryear {%
Samela%
\ \protect \BOthers {.}}{%
Samela%
\ \protect \BOthers {.}}{%
{\protect \APACyear {2016}}%
}]{%
samela2016based}
\APACinsertmetastar {%
samela2016based}%
\begin{APACrefauthors}%
Samela, C.%
, Manfreda, S.%
, Paola, F\BPBI D.%
, Giugni, M.%
, Sole, A.%
\BCBL {}\ \BBA {} Fiorentino, M.%
\end{APACrefauthors}%
\unskip\
\newblock
\APACrefYearMonthDay{2016}{}{}.
\newblock
{\BBOQ}\APACrefatitle {DEM-based approaches for the delineation of flood-prone
  areas in an ungauged basin in Africa} {Dem-based approaches for the
  delineation of flood-prone areas in an ungauged basin in africa}.{\BBCQ}
\newblock
\APACjournalVolNumPages{Journal of Hydrologic Engineering}{21}{2}{06015010}.
\PrintBackRefs{\CurrentBib}

\bibitem [\protect \citeauthoryear {%
Sangireddy%
, Stark%
, Kladzyk%
\BCBL {}\ \BBA {} Passalacqua%
}{%
Sangireddy%
\ \protect \BOthers {.}}{%
{\protect \APACyear {2016}}%
}]{%
sangireddy2016geonet}
\APACinsertmetastar {%
sangireddy2016geonet}%
\begin{APACrefauthors}%
Sangireddy, H.%
, Stark, C\BPBI P.%
, Kladzyk, A.%
\BCBL {}\ \BBA {} Passalacqua, P.%
\end{APACrefauthors}%
\unskip\
\newblock
\APACrefYearMonthDay{2016}{}{}.
\newblock
{\BBOQ}\APACrefatitle {GeoNet: An open source software for the automatic and
  objective extraction of channel heads, channel network, and channel
  morphology from high resolution topography data} {Geonet: An open source
  software for the automatic and objective extraction of channel heads, channel
  network, and channel morphology from high resolution topography data}.{\BBCQ}
\newblock
\APACjournalVolNumPages{Environmental Modelling \& Software}{83}{}{58--73}.
\PrintBackRefs{\CurrentBib}

\bibitem [\protect \citeauthoryear {%
Scriven%
, McGrath%
\BCBL {}\ \BBA {} Stefanakis%
}{%
Scriven%
\ \protect \BOthers {.}}{%
{\protect \APACyear {2021}}%
}]{%
scriven2021gis}
\APACinsertmetastar {%
scriven2021gis}%
\begin{APACrefauthors}%
Scriven, B\BPBI W\BPBI G.%
, McGrath, H.%
\BCBL {}\ \BBA {} Stefanakis, E.%
\end{APACrefauthors}%
\unskip\
\newblock
\APACrefYearMonthDay{2021}{}{}.
\newblock
{\BBOQ}\APACrefatitle {GIS derived synthetic rating curves and HAND model to
  support on-the-fly flood mapping} {Gis derived synthetic rating curves and
  hand model to support on-the-fly flood mapping}.{\BBCQ}
\newblock
\APACjournalVolNumPages{Natural Hazards}{109}{2}{1629--1653}.
\PrintBackRefs{\CurrentBib}

\bibitem [\protect \citeauthoryear {%
Seleem%
, Ayzel%
, Bronstert%
\BCBL {}\ \BBA {} Heistermann%
}{%
Seleem%
\ \protect \BOthers {.}}{%
{\protect \APACyear {2022}}%
}]{%
seleem2022transferability}
\APACinsertmetastar {%
seleem2022transferability}%
\begin{APACrefauthors}%
Seleem, O.%
, Ayzel, G.%
, Bronstert, A.%
\BCBL {}\ \BBA {} Heistermann, M.%
\end{APACrefauthors}%
\unskip\
\newblock
\APACrefYearMonthDay{2022}{}{}.
\newblock
{\BBOQ}\APACrefatitle {Transferability of data-driven models to predict urban
  pluvial flood water depth in Berlin, Germany} {Transferability of data-driven
  models to predict urban pluvial flood water depth in berlin, germany}.{\BBCQ}
\newblock
\APACjournalVolNumPages{Natural Hazards and Earth System Sciences
  Discussions}{2022}{}{1--23}.
\PrintBackRefs{\CurrentBib}

\bibitem [\protect \citeauthoryear {%
Tavares~da Costa%
\ \protect \BOthers {.}}{%
Tavares~da Costa%
\ \protect \BOthers {.}}{%
{\protect \APACyear {2020}}%
}]{%
tavares2020predictive}
\APACinsertmetastar {%
tavares2020predictive}%
\begin{APACrefauthors}%
Tavares~da Costa, R.%
, Zanardo, S.%
, Bagli, S.%
, Hilberts, A\BPBI G.%
, Manfreda, S.%
, Samela, C.%
\BCBL {}\ \BBA {} Castellarin, A.%
\end{APACrefauthors}%
\unskip\
\newblock
\APACrefYearMonthDay{2020}{}{}.
\newblock
{\BBOQ}\APACrefatitle {Predictive modeling of envelope flood extents using
  geomorphic and climatic-hydrologic catchment characteristics} {Predictive
  modeling of envelope flood extents using geomorphic and climatic-hydrologic
  catchment characteristics}.{\BBCQ}
\newblock
\APACjournalVolNumPages{Water Resources Research}{56}{9}{e2019WR026453}.
\PrintBackRefs{\CurrentBib}

\bibitem [\protect \citeauthoryear {%
{Technical Mapping Advisory Council}%
}{%
{Technical Mapping Advisory Council}%
}{%
{\protect \APACyear {2020}}%
}]{%
tmac2021flooding}
\APACinsertmetastar {%
tmac2021flooding}%
\begin{APACrefauthors}%
{Technical Mapping Advisory Council}.%
\end{APACrefauthors}%
\unskip\
\newblock
\APACrefYearMonthDay{2020}{}{}.
\newblock
\APACrefbtitle {Technical Mapping Advisory Council Annual Report 2020}
  {Technical mapping advisory council annual report 2020}\
  \APACbVolEdTR{}{\BTR{}}.
\newblock
\APACaddressInstitution{Washington, DC}{Federal Emergency Management Agency}.
\PrintBackRefs{\CurrentBib}

\bibitem [\protect \citeauthoryear {%
Thornton%
, Shrestha%
, Wei%
, Thornton%
\BCBL {}\ \BBA {} Kao%
}{%
Thornton%
\ \protect \BOthers {.}}{%
{\protect \APACyear {2022}}%
}]{%
thornton2022daymet}
\APACinsertmetastar {%
thornton2022daymet}%
\begin{APACrefauthors}%
Thornton, M.%
, Shrestha, R.%
, Wei, Y.%
, Thornton, P.%
\BCBL {}\ \BBA {} Kao, S\BHBI C.%
\end{APACrefauthors}%
\unskip\
\newblock
\APACrefYearMonthDay{2022}{}{}.
\newblock
\APACrefbtitle {Daymet: {D}aily {S}urface {W}eather {D}ata on a 1-km {G}rid for
  {N}orth {A}merica, {V}ersion 4 R1 [{D}ataset].} {Daymet: {D}aily {S}urface
  {W}eather {D}ata on a 1-km {G}rid for {N}orth {A}merica, {V}ersion 4 r1
  [{D}ataset].}
\newblock
\APACaddressPublisher{}{{ORNL DAAC, Oak Ridge, Tennessee, USA}}.
\newblock
\begin{APACrefURL} \url{https://doi.org/10.3334/ORNLDAAC/2129} \end{APACrefURL}
\newblock
\APACrefnote{Accessed: 2024-11-01}
\PrintBackRefs{\CurrentBib}

\bibitem [\protect \citeauthoryear {%
{U.S. Department of Agriculture}%
}{%
{U.S. Department of Agriculture}%
}{%
{\protect \APACyear {2024}}%
}]{%
SSURGO_Database}
\APACinsertmetastar {%
SSURGO_Database}%
\begin{APACrefauthors}%
{U.S. Department of Agriculture}.%
\end{APACrefauthors}%
\unskip\
\newblock
\APACrefYearMonthDay{2024}{}{}.
\newblock
\APACrefbtitle {Soil {S}urvey {G}eographic ({SSURGO}) {D}atabase [{D}ataset].}
  {Soil {S}urvey {G}eographic ({SSURGO}) {D}atabase [{D}ataset].}
\newblock
\APACaddressPublisher{}{Soil Survey Staff, Natural Resources Conservation
  Service}.
\newblock
\begin{APACrefURL} \url{https://nrcs.app.box.com/v/soils/folder/17971946225}
  \end{APACrefURL}
\newblock
\APACrefnote{Accessed: 2024-11-01}
\PrintBackRefs{\CurrentBib}

\bibitem [\protect \citeauthoryear {%
{U.S. Geological Survey}%
}{%
{U.S. Geological Survey}%
}{%
{\protect \APACyear {2024a}}%
}]{%
USGS_3DEP_DEM}
\APACinsertmetastar {%
USGS_3DEP_DEM}%
\begin{APACrefauthors}%
{U.S. Geological Survey}.%
\end{APACrefauthors}%
\unskip\
\newblock
\APACrefYearMonthDay{2024a}{}{}.
\newblock
\APACrefbtitle {{3D} {E}levation {P}rogram {D}igital {E}levation {M}odel
  [{D}ataset].} {{3D} {E}levation {P}rogram {D}igital {E}levation {M}odel
  [{D}ataset].}
\newblock
\begin{APACrefURL}
  \url{https://prd-tnm.s3.amazonaws.com/index.html?prefix=StagedProducts/Elevation/}
  \end{APACrefURL}
\newblock
\APACrefnote{Accessed: 2024-11-01}
\PrintBackRefs{\CurrentBib}

\bibitem [\protect \citeauthoryear {%
{U.S. Geological Survey}%
}{%
{U.S. Geological Survey}%
}{%
{\protect \APACyear {2024b}}%
}]{%
USGS_NHD}
\APACinsertmetastar {%
USGS_NHD}%
\begin{APACrefauthors}%
{U.S. Geological Survey}.%
\end{APACrefauthors}%
\unskip\
\newblock
\APACrefYearMonthDay{2024b}{}{}.
\newblock
\APACrefbtitle {National {H}ydrography {D}ataset ({NHD}) [{D}ataset].}
  {National {H}ydrography {D}ataset ({NHD}) [{D}ataset].}
\newblock
\begin{APACrefURL}
  \url{https://prd-tnm.s3.amazonaws.com/index.html?prefix=StagedProducts/Hydrography/}
  \end{APACrefURL}
\newblock
\APACrefnote{Accessed: 2024-11-01}
\PrintBackRefs{\CurrentBib}

\bibitem [\protect \citeauthoryear {%
{U.S. Geological Survey}%
}{%
{U.S. Geological Survey}%
}{%
{\protect \APACyear {2024c}}%
}]{%
USGS_2024_NLCD}
\APACinsertmetastar {%
USGS_2024_NLCD}%
\begin{APACrefauthors}%
{U.S. Geological Survey}.%
\end{APACrefauthors}%
\unskip\
\newblock
\APACrefYearMonthDay{2024c}{}{}.
\newblock
\APACrefbtitle {Annual {N}ational {L}and {C}over {D}atabase ({NLCD}) Collection
  1 Products [{D}ataset].} {Annual {N}ational {L}and {C}over {D}atabase
  ({NLCD}) collection 1 products [{D}ataset].}
\newblock
\APACaddressPublisher{}{Earth Resources Observation and Science (EROS) Center}.
\newblock
\APACrefnote{Accessed: 2024-11-01}
\newblock
\begin{APACrefDOI} \doi{10.5066/P94UXNTS} \end{APACrefDOI}
\PrintBackRefs{\CurrentBib}

\bibitem [\protect \citeauthoryear {%
Van Der~Knijff%
, Younis%
\BCBL {}\ \BBA {} De~Roo%
}{%
Van Der~Knijff%
\ \protect \BOthers {.}}{%
{\protect \APACyear {2010}}%
}]{%
van2010lisflood}
\APACinsertmetastar {%
van2010lisflood}%
\begin{APACrefauthors}%
Van Der~Knijff, J.%
, Younis, J.%
\BCBL {}\ \BBA {} De~Roo, A.%
\end{APACrefauthors}%
\unskip\
\newblock
\APACrefYearMonthDay{2010}{}{}.
\newblock
{\BBOQ}\APACrefatitle {{LISFLOOD}: a {GIS}-based distributed model for river
  basin scale water balance and flood simulation} {{LISFLOOD}: a {GIS}-based
  distributed model for river basin scale water balance and flood
  simulation}.{\BBCQ}
\newblock
\APACjournalVolNumPages{International Journal of Geographical Information
  Science}{24}{2}{189--212}.
\PrintBackRefs{\CurrentBib}

\bibitem [\protect \citeauthoryear {%
Wagenaar%
, L{\"u}dtke%
, Schr{\"o}ter%
, Bouwer%
\BCBL {}\ \BBA {} Kreibich%
}{%
Wagenaar%
\ \protect \BOthers {.}}{%
{\protect \APACyear {2018}}%
}]{%
wagenaar2018regional}
\APACinsertmetastar {%
wagenaar2018regional}%
\begin{APACrefauthors}%
Wagenaar, D.%
, L{\"u}dtke, S.%
, Schr{\"o}ter, K.%
, Bouwer, L\BPBI M.%
\BCBL {}\ \BBA {} Kreibich, H.%
\end{APACrefauthors}%
\unskip\
\newblock
\APACrefYearMonthDay{2018}{}{}.
\newblock
{\BBOQ}\APACrefatitle {Regional and temporal transferability of multivariable
  flood damage models} {Regional and temporal transferability of multivariable
  flood damage models}.{\BBCQ}
\newblock
\APACjournalVolNumPages{Water Resources Research}{54}{5}{3688--3703}.
\PrintBackRefs{\CurrentBib}

\bibitem [\protect \citeauthoryear {%
Wasko%
, Nathan%
, Stein%
\BCBL {}\ \BBA {} O'Shea%
}{%
Wasko%
\ \protect \BOthers {.}}{%
{\protect \APACyear {2021}}%
}]{%
wasko2021evidence}
\APACinsertmetastar {%
wasko2021evidence}%
\begin{APACrefauthors}%
Wasko, C.%
, Nathan, R.%
, Stein, L.%
\BCBL {}\ \BBA {} O'Shea, D.%
\end{APACrefauthors}%
\unskip\
\newblock
\APACrefYearMonthDay{2021}{}{}.
\newblock
{\BBOQ}\APACrefatitle {Evidence of shorter more extreme rainfalls and increased
  flood variability under climate change} {Evidence of shorter more extreme
  rainfalls and increased flood variability under climate change}.{\BBCQ}
\newblock
\APACjournalVolNumPages{Journal of Hydrology}{603}{}{126994}.
\PrintBackRefs{\CurrentBib}

\bibitem [\protect \citeauthoryear {%
Q.~Yang%
\ \protect \BOthers {.}}{%
Q.~Yang%
\ \protect \BOthers {.}}{%
{\protect \APACyear {2021}}%
}]{%
yang2021high}
\APACinsertmetastar {%
yang2021high}%
\begin{APACrefauthors}%
Yang, Q.%
, Shen, X.%
, Anagnostou, E\BPBI N.%
, Mo, C.%
, Eggleston, J\BPBI R.%
\BCBL {}\ \BBA {} Kettner, A\BPBI J.%
\end{APACrefauthors}%
\unskip\
\newblock
\APACrefYearMonthDay{2021}{}{}.
\newblock
{\BBOQ}\APACrefatitle {A high-resolution flood inundation archive (2016--the
  present) from Sentinel-1 SAR imagery over CONUS} {A high-resolution flood
  inundation archive (2016--the present) from sentinel-1 sar imagery over
  conus}.{\BBCQ}
\newblock
\APACjournalVolNumPages{Bulletin of the American Meteorological
  Society}{102}{5}{E1064--E1079}.
\PrintBackRefs{\CurrentBib}

\bibitem [\protect \citeauthoryear {%
Z.~Yang%
, Bai%
\BCBL {}\ \BBA {} Xiang%
}{%
Z.~Yang%
\ \protect \BOthers {.}}{%
{\protect \APACyear {2019}}%
}]{%
yang2019lattice}
\APACinsertmetastar {%
yang2019lattice}%
\begin{APACrefauthors}%
Yang, Z.%
, Bai, F.%
\BCBL {}\ \BBA {} Xiang, K.%
\end{APACrefauthors}%
\unskip\
\newblock
\APACrefYearMonthDay{2019}{}{}.
\newblock
{\BBOQ}\APACrefatitle {A lattice Boltzmann model for the open channel flows
  described by the Saint-Venant equations} {A lattice boltzmann model for the
  open channel flows described by the saint-venant equations}.{\BBCQ}
\newblock
\APACjournalVolNumPages{Royal Society open science}{6}{11}{190439}.
\PrintBackRefs{\CurrentBib}

\bibitem [\protect \citeauthoryear {%
Zheng%
, Maidment%
, Tarboton%
, Liu%
\BCBL {}\ \BBA {} Passalacqua%
}{%
Zheng%
, Maidment%
\BCBL {}\ \protect \BOthers {.}}{%
{\protect \APACyear {2018}}%
}]{%
zheng2018geoflood}
\APACinsertmetastar {%
zheng2018geoflood}%
\begin{APACrefauthors}%
Zheng, X.%
, Maidment, D\BPBI R.%
, Tarboton, D\BPBI G.%
, Liu, Y\BPBI Y.%
\BCBL {}\ \BBA {} Passalacqua, P.%
\end{APACrefauthors}%
\unskip\
\newblock
\APACrefYearMonthDay{2018}{}{}.
\newblock
{\BBOQ}\APACrefatitle {GeoFlood: Large-scale flood inundation mapping based on
  high-resolution terrain analysis} {Geoflood: Large-scale flood inundation
  mapping based on high-resolution terrain analysis}.{\BBCQ}
\newblock
\APACjournalVolNumPages{Water Resources Research}{54}{12}{10--013}.
\PrintBackRefs{\CurrentBib}

\bibitem [\protect \citeauthoryear {%
Zheng%
, Tarboton%
, Maidment%
, Liu%
\BCBL {}\ \BBA {} Passalacqua%
}{%
Zheng%
, Tarboton%
\BCBL {}\ \protect \BOthers {.}}{%
{\protect \APACyear {2018}}%
}]{%
zheng2018river}
\APACinsertmetastar {%
zheng2018river}%
\begin{APACrefauthors}%
Zheng, X.%
, Tarboton, D\BPBI G.%
, Maidment, D\BPBI R.%
, Liu, Y\BPBI Y.%
\BCBL {}\ \BBA {} Passalacqua, P.%
\end{APACrefauthors}%
\unskip\
\newblock
\APACrefYearMonthDay{2018}{}{}.
\newblock
{\BBOQ}\APACrefatitle {River channel geometry and rating curve estimation using
  height above the nearest drainage} {River channel geometry and rating curve
  estimation using height above the nearest drainage}.{\BBCQ}
\newblock
\APACjournalVolNumPages{JAWRA Journal of the American Water Resources
  Association}{54}{4}{785--806}.
\PrintBackRefs{\CurrentBib}

\end{thebibliography}

\end{document}